\documentclass{WileyMSP-template}

\usepackage{subcaption}
\usepackage{svg}
\usepackage{amsmath}
\usepackage{amssymb}
\usepackage{amsfonts}
\usepackage{color}

\begin{document}

\pagestyle{fancy}
\rhead{\includegraphics[width=2.5cm]{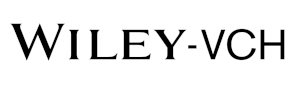}}

\title{Real-Time Feedback-Driven Single-Particle Tracking: A Survey and Perspective}

\maketitle

% Author: Please give full first and last names for authors and include * after the name of all corresponding authors
\author{Bertus van Heerden}%$^{1,2,\dagger}$\orcidB{}}
\author{Nicholas A. Vickers}% $^{3,\dagger}$\orcidA{}}
\author{Tjaart P.J. Krüger$^{*}$}% $^{1,2,}$*\orcidC{}}
\author{Sean B. Andersson$^{*}$}%$^{3,4,}$* \orcidD{}}

% Dedication
\dedication{}

% Affiliations: Please provide adacemic titles (Prof. or Dr.) for all authors where applicable, and include an institutional email address for all corresponding authors
\begin{affiliations}
B. van Heerden$^{1,2}$, Prof. T.P.J. Krüger$^{1,2}$\\
$^{1}$ \quad Department of Physics, University of Pretoria, %Lynnwood Road, 
Pretoria, 0002, South Africa\\
$^{2}$ \quad Forestry and Agricultural Biotechnology Institute (FABI), University of Pretoria, Pretoria, South Africa\\
Email Address: bertus.vanheerden@up.ac.za (B.v.H.), tjaart.kruger@up.ac.za (T.P.J.K.)\\

N.A. Vickers$^{3}$, Prof. S.B. Andersson$^{3,4}$\\
$^{3}$ \quad Department of Mechanical Engineering, Boston University, Boston, MA, USA\\
$^{4}$ \quad Division of Systems Engineering, Boston University, Boston, MA, USA\\
Email Address: nvickers@bu.edu (N.A.V.), sanderss@bu.edu (S.B.A.)
\end{affiliations}

% Keywords: Please provide a minimum of three and a maximum of seven keywords, separated by commas

% Abstract should be written in the present tense and impersonal style (i.e., avoid we), and be at most 200 words long
\begin{abstract}
\justifying
Real-time feedback-driven single-particle tracking (RT-FD-SPT) is a class of techniques in the field of single-particle tracking that uses feedback control to keep a particle of interest in a detection volume. These methods provide high spatiotemporal resolution on particle dynamics and allow for concurrent spectroscopic measurements. This review article begins with a survey of existing techniques and of applications where RT-FD-SPT has played an important role. We then systematically discuss each of the core components of RT-FD-SPT in order to develop an understanding of the trade-offs that must be made in algorithm design and to create a clear picture of the important differences, advantages, and drawbacks of existing approaches. These components are feedback tracking and control, ranging from simple proportional-integral-derivative control to advanced nonlinear techniques, estimation to determine particle location from the measured data, including both online and offline algorithms, and techniques for calibrating and characterizing different RT-FD-SPT methods. We then introduce a collection of metrics for RT-FD-SPT to help guide experimentalists in selecting a method for their particular application and to help reveal where there are gaps in the techniques that represent opportunities for further development. Finally, we conclude with a discussion on future perspectives in the field.
\end{abstract}

% Text: Please use section headings and subheadings as specified below. For communications, all section headings apart from Experimental Section should be removed
% Please make the first reference to a display item bold: \textbf{Figure 1}
% Do not abbreviate Figure, Equation, etc.; display items are always singular, i.e., Figure 1 and 2.
% Equations are always singular, i.e., Equation 1 and 2, and should be inserted using the {equation} environment, not as graphics
% Please do not use footnotes in the text, additional information can be added to the Reference list.

\section{Introduction}\label{sec:Introduction}
% goal describe and motivate why one would want to track particles and why RT-FD-SPT is important and valueable
 %The field of optical microscopy is constantly advancing spatiotemporal resolution. State-of-the-art techniques allow the visualization of single molecules through localization with nanometer spatial resolution at a time resolution of 1 ms or better.
 \justifying
The ability to detect individual molecules and observe their dynamical behavior has revolutionized life sciences during the past three decades. State-of-the-art techniques provide nanometer-scale spatial resolution at a time resolution of 1 ms or better. An important single-molecule microscopy technique, known as single-particle tracking (SPT)~\cite{Dupont2011,Chenouard2014,Manzo2015,Liu2015a,Mathai2016,Shen2017,VonDiezmann2017,Zhong2020}, allows direct access to the dynamics of individual freely diffusing molecules and particles in live cells.  The conventional approach to SPT involves acquiring a sequence of images of fluorescent particles, determining their positions in each frame, and linking those into a trajectory for further analysis. Such image-based SPT can achieve excellent three-dimensional spatial resolution (10--20 nm)~\cite{Thompson2010,Zhong2017,Wang2017}, but the time resolution is limited to 30--100 ms~\cite{Zhong2017,Bouchard2015,Wang2017} due to the integration time needed to collect enough photons (ca. 500--10,000)~\cite{Wang2017,Thompson2010} to form an image.

Another single-molecule detection technique that has significantly advanced our understanding of the properties and dynamics of biomolecules is known as single-molecule spectroscopy (SMS)~\cite{Moerner2002}. SMS is an umbrella term for various types of measurements on single molecules and molecular or nanoscale systems that involve photoexcitation of photoactive labels attached to the molecular systems or direct excitation of the molecules or particles. Fluorescence photons are typically detected and used to obtain information such as the emission brightness, spectra, and fluorescence lifetimes. The selectivity and sensitivity are significantly enhanced by isolating the biomolecules from their natural environment, and the information content is increased by a few orders of magnitude through immobilization of the molecules or confinement of their motion to a small subspace to allow sufficiently long observation times. Such modifications, however, significantly alter the setting in which the target molecules are being studied. An important challenge facing SMS, therefore, is the ability to measure biological molecules in their natural environment and for sufficiently long durations to obtain meaningful information but without compromising their structure and dynamics.

Real-time feedback-driven single-particle tracking (RT-FD-SPT) is a technique that offers a marked improvement in temporal resolution and an extended range-- especially in the axial direction--  compared to image-based SPT and additionally enables concurrent spectroscopic measurements to be made on the tracked particles. In most cases, image-based SPT does not allow concurrent SMS because the calculation of the particle location is too slow. In contrast, RT-FD-SPT allows SMS to be performed on freely moving biomolecules. This capability overcomes some of the major current limitations of standard SMS, allowing for long observation times without significant alteration of the biomolecular motion. RT-FD-SPT thus represents an important stepping stone towards realizing \textit{in vivo} SMS.

Thus far, the development of RT-FD-SPT has been somewhat 
fractured, with different research groups coming up with independent solutions. This is due, at least in part, to its interdisciplinary nature, with approaches relying on techniques from physics, engineering, chemistry, and biology. There have thus far been only three reviews focusing solely on RT-FD-SPT~\cite{Cang2008,Welsher2015,Hou2019a}. While some reviews of image-based SPT have touched upon RT-FD-SPT methods~\cite{Dupont2011,Mathai2016,Shen2017,Zhong2020}, they do not consider this topic in depth. %(some general reviews of SPT have addressed image-based and RT-FD-SPT approaches~\cite{Dupont2011,Mathai2016,Shen2017,Zhong2020}). 
Ref.~\cite{Cang2008} dates from 2008 and there have since been many developments. Ref.~\cite{Welsher2015} focused almost exclusively on one specific technique (3D multi-resolution microscopy). The most recent review~\cite{Hou2019a} described a variety of methods and compared their experimental performance but did not present a systematic analysis of RT-FD-SPT. Because the different approaches towards RT-FD-SPT have different strengths and weaknesses, it can be difficult to decide on a particular method for a particular application. Moreover, in most published work the parameters underlying the algorithms are typically hand-tuned until satisfactory performance is achieved, and the methods are presented without a clear picture of what the important differences, advantages, drawbacks, and trade-offs are. 
%illustrates the current state of the art. The different methods were described and their experimental performance was compared, but the review did not present a systematic analysis of the topic. This issue is seen in much of the research thus far, where different methods are often presented without a clear picture of what the important differences, advantages, and drawbacks are.

This work aims to more systematically approach the topic, discussing each component of the tracking system and how it affects performance.  We hope that this effort will help develop a more unified view of RT-FD-SPT, making it more accessible to a wide range of scientists. The review also aims to be more complete than the previous three reviews, since especially techniques focusing on improved feedback control and position estimation have been overlooked thus far. Image-based SPT has been comprehensively reviewed, but the same cannot be said for RT-FD-SPT. We address control theory and parameter estimation, two aspects of RT-FD-SPT that are generally only cursorily considered by most research in the field, and look at the criteria used to evaluate performance when choosing a method. Lastly, we aim to highlight the potential of concurrent spectroscopy, which has also been under-emphasized in the research to date.

The review is structured as follows: \textbf{Section~\ref{sec:Survey}} presents a survey of RT-FD-SPT techniques, while \textbf{Section~\ref{sec:Applications}} discusses applications in microscopy and spectroscopy. \textbf{Sections~\ref{sec:Control} and \ref{sec:estimation}} respectively discuss the aspects of control and estimation theory relevant to RT-FD-SPT, while \textbf{Section~\ref{sec:calibration}} discusses system characterization and calibration. Finally, \textbf{Section~\ref{sec:Performance}} discusses the performance of the different methods and \textbf{Section~\ref{sec:outlook}} presents an outlook to the future of the field.

\section{Survey} \label{sec:Survey}
%description of the types of SPT hardware

\begin{figure}[t!]
    \centering
    \includegraphics[width=14 cm]{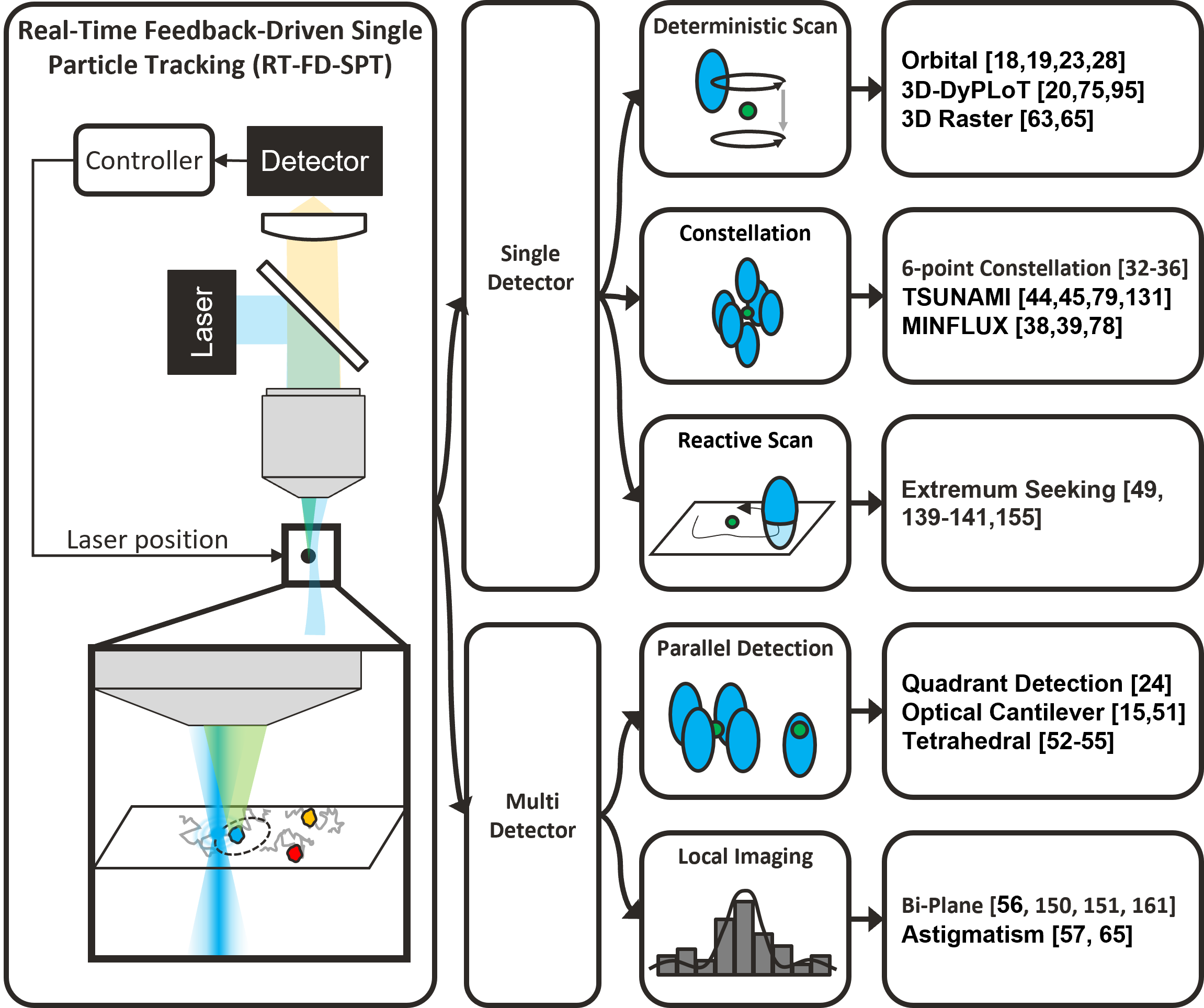}
    \caption{A taxonomy of current 3D RT-FD-SPT techniques. The class of SPT microscopes that use feedback control are depicted on the far left. These microscopes use information gained from photon detections to control the optics and lasers enabling it to lock onto a moving particle or molecule. This class can be grouped by the number of detectors used.  Methods that use a single photodetector can be further grouped based on how the laser is scanned. Three groupings of scanning methods emerge: deterministic scan, constellation scanning pattern, and reactive scan. Deterministic scanning moves the laser along a predetermined path; depicted is the 3D-orbital method. The results are then used to adjust the location of the center of the scan using feedback. Constellation methods dwell the laser at a set of predetermined locations; depicted is a typical constellation. The results are used to adjust the location of the center of the pattern using feedback. Reactive scans determine the next direction of the laser through feedback determined by the most recent photon detections; depicted is the extremum seeking method. The group of methods using multiple photodetectors can be further broken down into methods that use a small number of detectors, and those that use cameras. Parallel detection methods utilize multiple photodetectors to determine the $x$, $y$, and $z$ positions simultaneously; depicted is the optical cantilever technique, showing the detection pattern for determining $x$ and $y$ positions on the left and the channel for determining the $z$ position on the right. Local imaging methods use a reduced region of interest on a scientific imaging camera; depicted are the photon counts of a row of pixels of the bi-plane method.}
    \label{fig:taxonomy}
\end{figure}

RT-FD-SPT differs from conventional image-based SPT in that the particle in question is tracked in real time, with a feedback loop keeping the particle in the observation volume. This offers several advantages. Firstly, particles can be tracked over long distances in all three dimensions, while maintaining excellent spatiotemporal resolution (the temporal resolution is routinely $\sim\!1$ ms or less). Secondly, the particle can explore a variety of environments, whereas conventional SPT is typically limited to a single region of interest. This is due in part to the long range capability (typically $\sim\! 10$ \textmu m, with a range from 50 \textmu m to a few centimeters~\cite{Wehnekamp2019}) of RT-FD-SPT and to the fact that the tracking parameters can be changed in real time to adapt to different environmental circumstances. Lastly, keeping the particle in a small observation volume allows concurrent single-molecule spectroscopic measurements to be made. The main drawback of RT-FD-SPT over conventional SPT is that only one particle can be tracked at a time, except when diffusion is slow enough that multiplexing is possible. There is also the possibility of losing the particle prematurely. Furthermore, RT-FD-SPT does not, by itself, provide contextual imaging. This can be provided by concurrent techniques (see \textbf{Section~\ref{sec:Applications}}), but this adds cost and complexity. Of course, such concurrent imaging may well allow conventional SPT, and in this way, the two techniques are complementary and can even be used in the same experiment. \textbf{Figure~\ref{fig:taxonomy}} shows a taxonomy of current RT-FD-SPT techniques, which can be grouped into single-detector and multi-detector methods. 

\begin{figure}[t]
\centering
    \begin{subfigure}[t]{0.9\textwidth}
        \includegraphics[width=\textwidth]{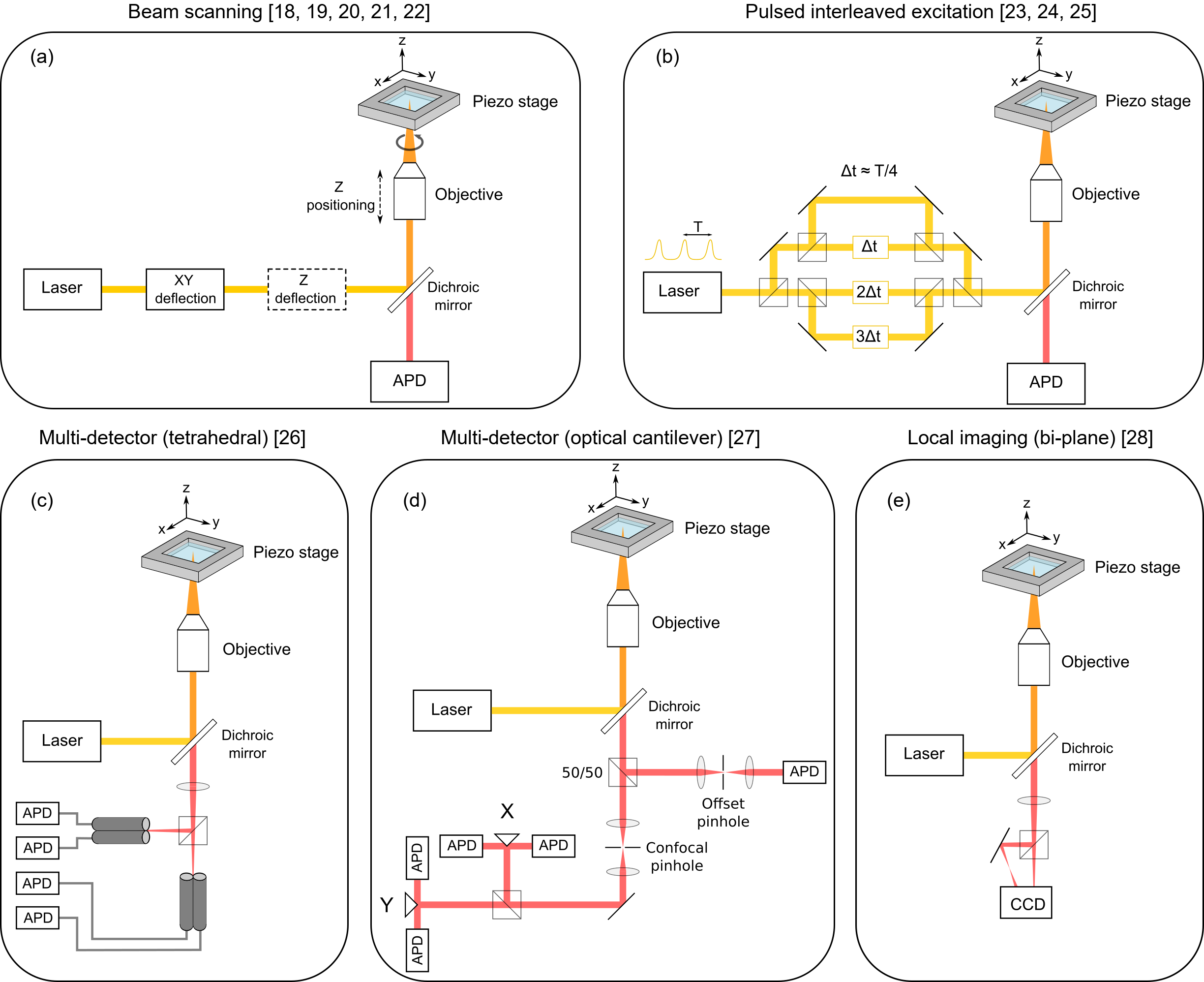}
        %\caption{Beam scanning~\cite{Gratton2005,Berglund2005,Hou2017,Shen2011,Balzarotti2017}.}
        \phantomcaption
        \label{fig:scanning}
    \end{subfigure}
    \begin{subfigure}[t]{0\textwidth}
        \includegraphics[width=\textwidth]{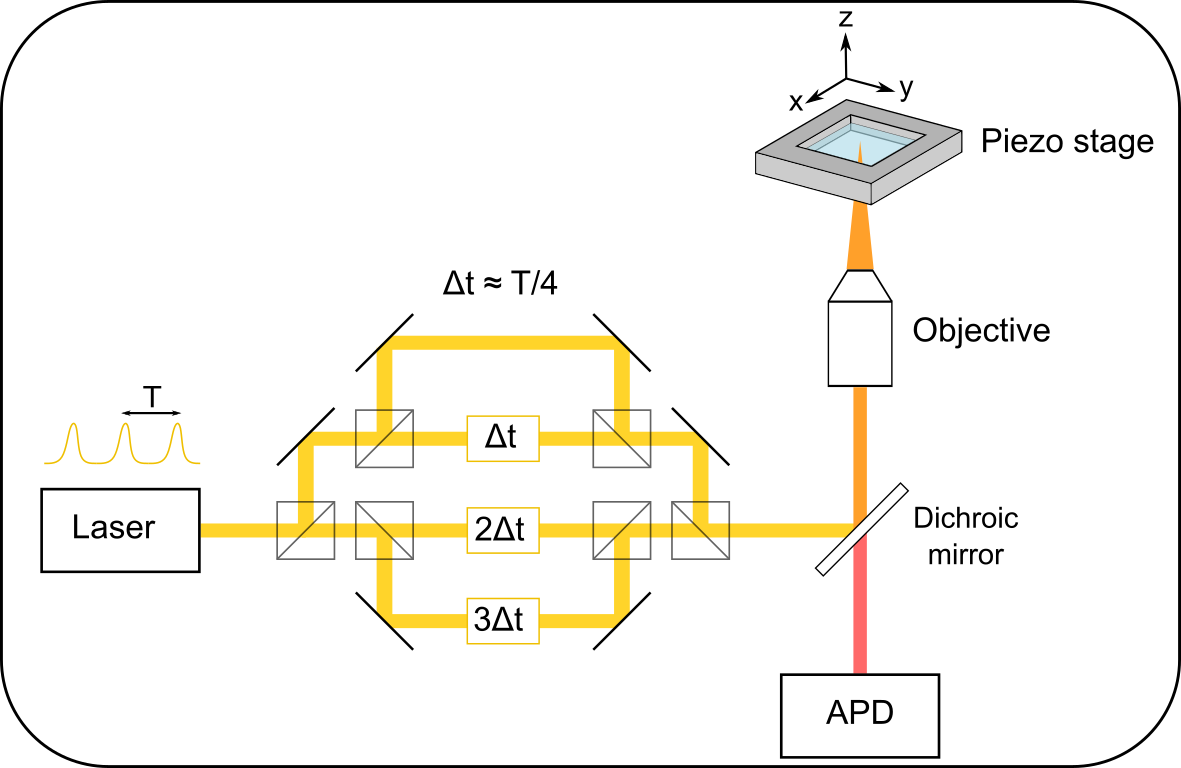}
        %\caption{Pulsed interleaved excitation~\cite{Davis2008,Perillo2015,Masullo2021}.}
        \phantomcaption
        \label{fig:pie}
    \end{subfigure}\vspace{8pt}
    \begin{subfigure}[t]{0\textwidth}
        \includegraphics[width=\textwidth]{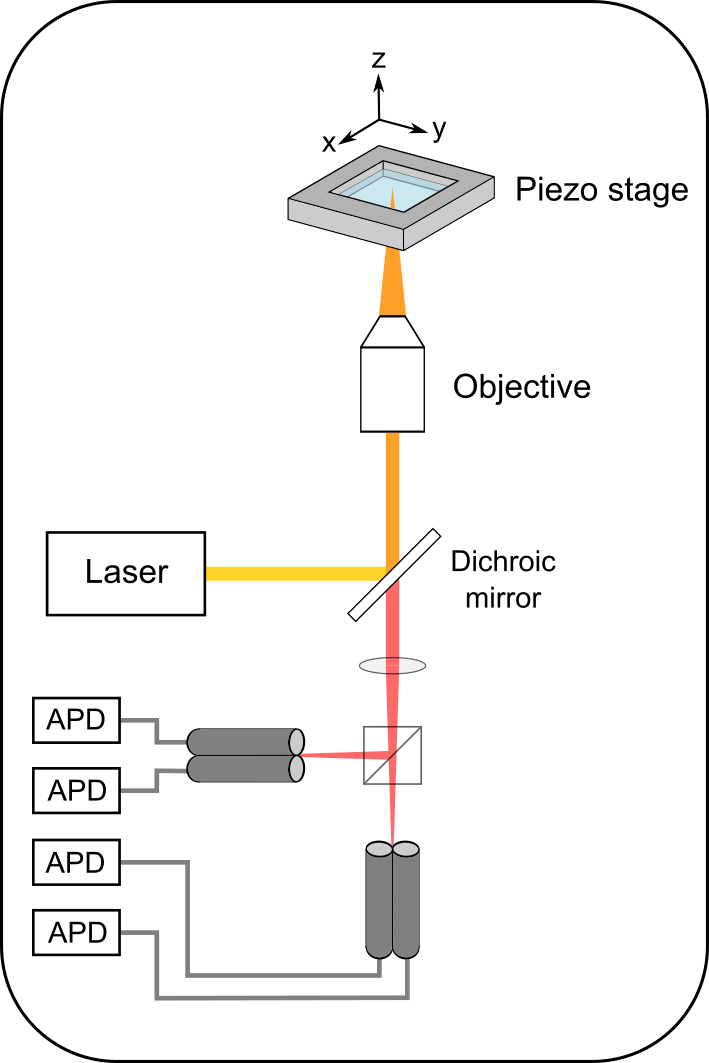}
        %\caption{Multi-detector (tetrahedral)~\cite{Lessard2006}.}
        \phantomcaption
        \label{fig:tetrahedral}
    \end{subfigure}
    \begin{subfigure}[t]{0\textwidth}
        \includegraphics[width=\textwidth]{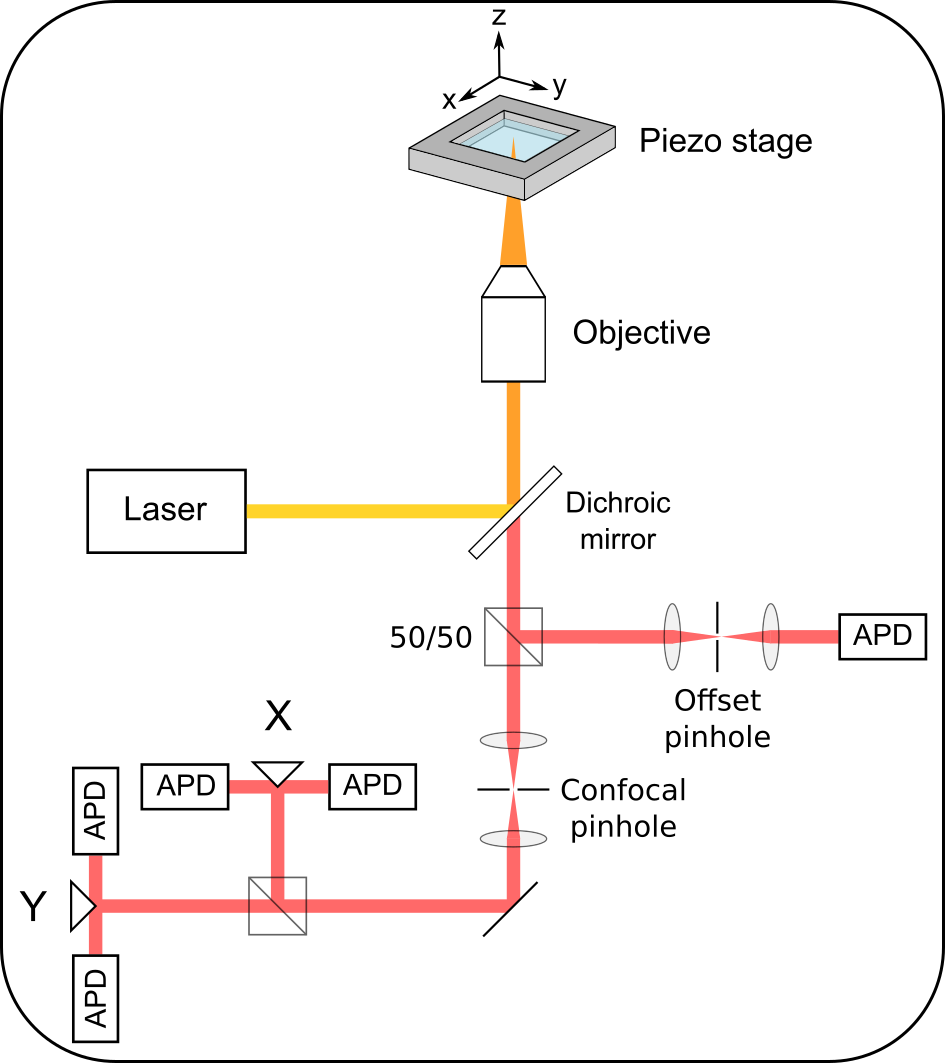}
        %\caption{Multi-detector (optical cantilever)~\cite{Cang2007}.}
        \phantomcaption
        \label{fig:cantilever}
    \end{subfigure}
    \begin{subfigure}[t]{0\textwidth}
        \includegraphics[width=\textwidth]{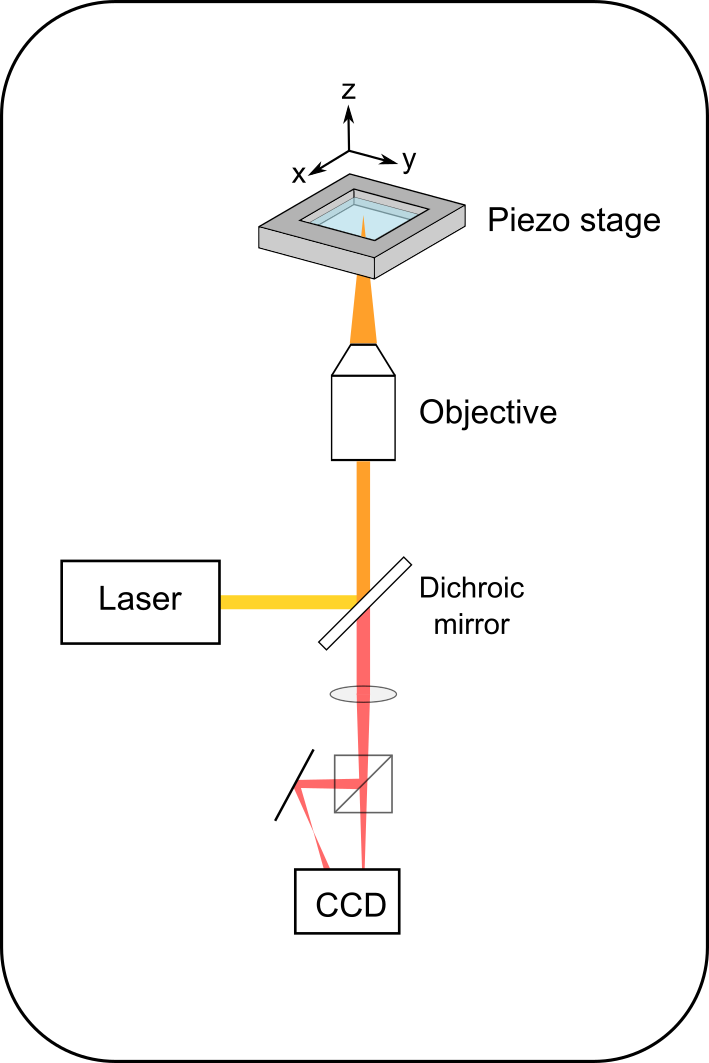}
        %\caption{Local imaging (bi-plane)~\cite{Juette2010}.}
        \phantomcaption
        \label{fig:localimage}
    \end{subfigure}
    \caption{Diagrams of the most common experimental configurations for RT-FD-SPT. In (a) the dashed lines indicate two different ways of including z-scanning, if used. Lasers can generally be continuous-wave or pulsed, except for the setup in (b), which requires a pulsed laser, as indicated. Avalanche photodiodes (APDs), specifically single-photon-sensitive APDs (SPADs), are usually implemented in the form of single-photon counting modules, and high-numerical aperture oil- or water-immersion objectives are typically used.}
\end{figure}

\subsection{Single-detector methods}
With a single detector, the position is determined by using time-varying excitation. The beam-scanning family of techniques (\textbf{Figure~\ref{fig:scanning}}) are based on fast scanning of the excitation beam, using piezoelectric or galvanometer mirrors, or with electro-optical or acousto-optical deflectors (EODs/AODs), with the latter two options being faster. 
Axial scanning can be performed by moving the objective~\cite{Gratton2005}, by using an electrically tunable lens (ETL)~\cite{Annibale2015}, or by using a tunable acoustic index gradient (TAG) lens~\cite{Hou2017}, with the speed increasing in the same order. Another way to obtain axial information is to switch between two focal planes using two beams that are switched on and off exactly out-of-phase using acousto-optic modulators (AOMs)~\cite{McHale2007}. It is also possible to scan the sample instead of the beam using a two- or three-dimensional piezo stage~\cite{Andersson2005,Lanzano2014}. This is conceptually equivalent, but has some practical differences. Scanning the sample with a stage is typically slower than scanning with a fast mirror or EOD/AOD. There is also the common concern that fluid dynamics in the sample due to the movement of the entire sample interferes with the particle motion. While this has been shown not to be a problem in practice~\cite{Cang2006}, beam scanning remains more popular for fast scanning.

The orbital scanning method was initially proposed by Enderlein~\cite{Enderlein2000a}, and first implemented independently by the groups of Gratton~\cite{Gratton2005} and Mabuchi~\cite{Berglund2005}. It was also used in one of the early versions of the Anti-Brownian ELectrokinetic (ABEL) trap~\cite{Cohen2008} (see \textbf{Section~\ref{sec:abel}} for more about the ABEL trap). This method involves scanning the beam in a circle around the particle and measuring the fluorescence signal with a point detector such as a single-photon avalanche diode (SPAD). The position of the particle can be inferred from the resulting sinusoidal temporal variation in the intensity signal. An alternative to orbital scanning is ``local raster scanning". This was first used for Levi and coworkers' ``Scanning fluorescence correlation spectroscopy (Scanning-FCS)"~\cite{Gratton2003}. 
Wang and Moerner~\cite{Wang2010,Wang2011} implemented this method using a knight's tour pattern in the ABEL trap and this approach was extended to 3D by Hou \textit{et al.}~\cite{Hou2017} to create 3D Dynamic Photon Localization Tracking (3D-DyPLoT). The advantage of this method is that it enables a larger tracking area, allowing tracking of faster particles. However, this comes with a trade-off of worse precision~\cite{VanHeerden2021}. Another disadvantage of the Knight's Tour method is that it involves discontinuous scanning. Photons detected \textit{between} scan points are not used for position estimation, and thus the scanning hardware needs to be very fast in order to spend as little time as possible moving between points. This adds significantly to the cost of the hardware. 

An alternative to a fine-grained scan like the orbital or local raster scan methods is to scan the beam through a constellation of a small number of points, as first implemented by Shen and Andersson~\cite{Shen2009,Shen2009a,Shen2011}, using a six-point constellation in 3D~\cite{Shen2011a}. This also requires fast scanning hardware, but has the advantage that the measurement locations can be chosen so as to maximize the position information gained from the observed photons~\cite{Shen2012,Zhang2021}. The same approach is taken with the superresolution technique MINFLUX~\cite{Balzarotti2017,Gwosch2020}, and a ``hybrid" approach was presented by Zhang and Welsher where a small number of XY points were used while scanning continuously in the Z direction~\cite{Zhang2021}.

An alternative to scanning the laser beam or sample stage is to use pulsed interleaved excitation (PIE) (\textbf{Figure~\ref{fig:pie}}). It was originally developed for techniques such as FCS and F\"orster resonance energy transfer (FRET)~\cite{Muller2005}. The use of PIE for RT-FD-SPT was first suggested by Dertinger \textit{et al.}~\cite{Enderlein2005}, and experimentally demonstrated by Davis \textit{et al.}~\cite{Davis2008,Gremann2014}, by using four alternately pulsed laser diodes with foci placed in a tetrahedral shape around the particle. The arrival times of the detected photons are then used to determine the position of the particle. Perillo \textit{et al.}~\cite{Perillo2015} implemented the idea in their microscope - named ``tracking single particles using nonlinear and multiplexed illumination (TSUNAMI)" by using a femtosecond laser beam that is split in four and sent along physical delay lines before being recombined. For time-gating the photon arrivals, a time-correlated single photon counting (TCSPC) system was used, as the pulses in different channels were only 3.3 ns apart. The technique was recently expanded to incorporate two-color detection~\cite{Liu2020a}. A pulsed interleaved version of MINFLUX was also recently presented~\cite{Masullo2021}, and PIE has been used for a three-dimensional version of the ABEL trap~\cite{Dissanayaka2019}.

Instead of scanning along a deterministic pattern or using PIE, it is also possible to use a control system that generates the scan path dynamically. This reactive scan method was first implemented by the Andersson group in the form of an extremum seeking controller~\cite{Andersson2011,Ashley2016a} that maximizes the detected intensity. Recently, Vickers and Andersson presented a controller design that aims to maximize the positional information gained from the measurement~\cite{Vickers2021}.

\subsection{Multi-detector methods}
Detection path-based methods can be based either on parallel detection or on local imaging. Parallel detection tracking was first developed by the groups of Yang~\cite{Cang2006,Cang2007}  and Werner~ \cite{Lessard2006,Lessard2007,DeVore2015a,Dunlap2020}. The approach taken by the two groups differs somewhat. The Yang group initially used a quadrant detector~\cite{Cang2006}, but later developed the so-called optical cantilever~\cite{Cang2007} (\textbf{Figure~\ref{fig:cantilever}}). The latter method made use of two detectors each for X and Y, as well as a detector for Z. The detected signal is split 50/50 between the XY and Z paths. The XY signal is split 50/50 again between X and Y and projected onto the centers of two prism mirrors with two detectors each. When the particle is centered on an axis, the signal is equal in both detectors for that axis. When the particle moves, the relative increase of the signal in one detector on an axis indicates the direction of motion. For axial position information, the signal is focused through a slightly offset pinhole onto the Z-detector. This gives a linear relationship between the detected intensity and the Z-position. The Werner group's version uses two pairs of detectors coupled to optical fibers, with the other ends of the fibers placed at two different focal planes, so as to correspond to a tetrahedron in the sample volume (\textbf{Figure~\ref{fig:tetrahedral}})~\cite{Lessard2006, Lessard2007, DeVore2015a, Dunlap2020}.

Local imaging refers to the use of a camera, such as a CCD or scientific CMOS camera, to image the particle and determine its position. The bi-plane method was developed by Juette and Bewersdorf~\cite{Juette2010}. They used two image planes in order to achieve 3D-localization, using different regions of an EMCCD camera to image both planes at the same time (\textbf{Figure~\ref{fig:localimage}}). By using only a $5\times5$ pixel region for each image, the readout and analysis of the image were computationally fast. Another imaging method that made use of real-time feedback was developed by Spille \textit{et al.}~\cite{Spille2015}. This method combined light sheet illumination with astigmatic detection to determine the axial position. A Z-axis piezo stage was used to refocus the sample in real time.

\subsection{Complementary methods}
\label{sec:abel}

There are many methods that share some features with RT-FD-SPT. One is wide-field image-based tracking, which is a well established and widely used technique. Another is the ABEL trap, which is sometimes (erroneously in our opinion) considered to be an SPT technique. It makes use of exactly the same type of position measurement and feedback, except that electric forces are used to reposition a particle in solution, instead of moving the sample or illumination beam. This allows for a much faster response time (particles with a diffusion coefficient $D \approx 100\ \mu $m$^2$ s$^{-1}$ have been trapped~\cite{Wang2011}). However, the particle needs to be isolated in solution, and this technique can thus not be used \textit{in vivo}. It can, however, be used to measure diffusion and electrokinetic mobility~\cite{Wang2011}, and for performing spectroscopy on particles in solution~\cite{Wang2012}. There are many other related trapping techniques, based on physical, optical or thermodynamic principles~\cite{Bespalova2019}. There are also techniques that use confocal detection to determine particle trajectories, but without using feedback. These include a multi-detector method similar to multi-detector RT-FD-SPT~\cite{Sahl2010,Sahl2014}, and a method that uses a single detector paired with a statistical method to extract the most likely trajectory from the photon counts~\cite{Jazani2019}.

\section{Applications}\label{sec:Applications}

Far more numerous than the approaches to RT-FD-SPT are the possible applications. RT-FD-SPT has been applied in many different biological contexts and its utility is mainly twofold. Firstly, it allows the observation of particle motion in three dimensions, at a high temporal rate and with an excellent signal-to-background ratio. To this end, the experimental setup is built around a microscope and is comparable to setups used for conventional SPT. This is usually combined with other, complementary imaging modalities. Secondly, spectroscopic measurements can be made on tracked particles. Both of these factors have proven to be very useful and we discuss them separately here.

\subsection{Microscopy and direct tracking-based applications}

In one of the first examples of feedback-based SPT, So \textit{et al.}~\cite{So1997SPT} used two-photon excitation to study phagocytosis, by tracking 2 $\mu$m latex spheres as they were captured by macrophages. Peters \textit{et al.}~\cite{Peters1998SPT} used a setup based on the deflection of a focused laser beam by a polystyrene sphere to measure motion of LFA-1 molecules in the cell membrane. Soon after the first demonstrations of the orbital tracking method, it was used to study phagocytosis of fluorescent beads by fibroblasts~\cite{Gratton2005}. It has since been used to study properties such as cell mechanics~\cite{Ragan2006,Wehnekamp2019}, membrane dynamics~\cite{Hellriegel2009}, transport~\cite{Katayama2009,Reuel2012,Dupont:2013bu,Lanzano2014,Verdeny-Vilanova2017,Annibale2015}, and DNA transcription~\cite{Donovan2019, marklund2020dna}. This is often done alongside imaging techniques such as light microscopy, fluorescence microscopy, two-photon microscopy, or superresolution microscopy.
The method has also been used to track nanoscaled structures such as changes in the shape of microvilli~\cite{Lanzano2011}, the dynamics of endolysosomes along microtubules~\cite{Verdeny-Vilanova2017}, and for tracking entire mitochondria in zebrafish larvae~\cite{Wehnekamp2019, Mieskes2020}. The Knight's Tour scanning pattern, originally implemented by Wang and Moerner~\cite{Wang2010} and extended to 3D by the Welsher group, was used to track virus-like particles landing on the cell membrane~\cite{Hou2019} and cellular trafficking of mRNA-encapsulating lipid nanoparticles~\cite{Patel2020}. MINFLUX has been used to study the diffusion of a ribosomal subunit in \textit{E. coli}~\cite{Balzarotti2017} and to image the Nup96 protein in the nuclear pore complex~\cite{Gwosch2020,Schmidt2021}, the post synaptic protein in neurons, and the MICOS complex in the mitochondrial inner membrane~\cite{Pape2020}. 
Tetrahedral excitation tracking has been used to study endocytosis and subcellular trafficking~\cite{Perillo2015,Liu2016} as well as the landing of nanoparticles on the plasma membrane~\cite{Liu2020a}. This technique was also combined with two-photon laser scanning microscopy (2P-LSM) in a very similar manner as for 3D multi-resolution microscopy (3D-MM). Multidetector tracking has been used for tracking membrane receptors~\cite{Wells2009}, including with simultaneous spinning-disk imaging~\cite{DeVore2015}. 3D-MM combined multi-detector tracking with 2P-LSM to achieve a detailed view of the cellular uptake process~\cite{Welsher2014}. %~(\textbf{Figure~\ref{fig:microscopy_mrm}}). 

\begin{figure}
    \centering
    \begin{subfigure}[t]{0.8\textwidth}
        \includegraphics[width=\textwidth]{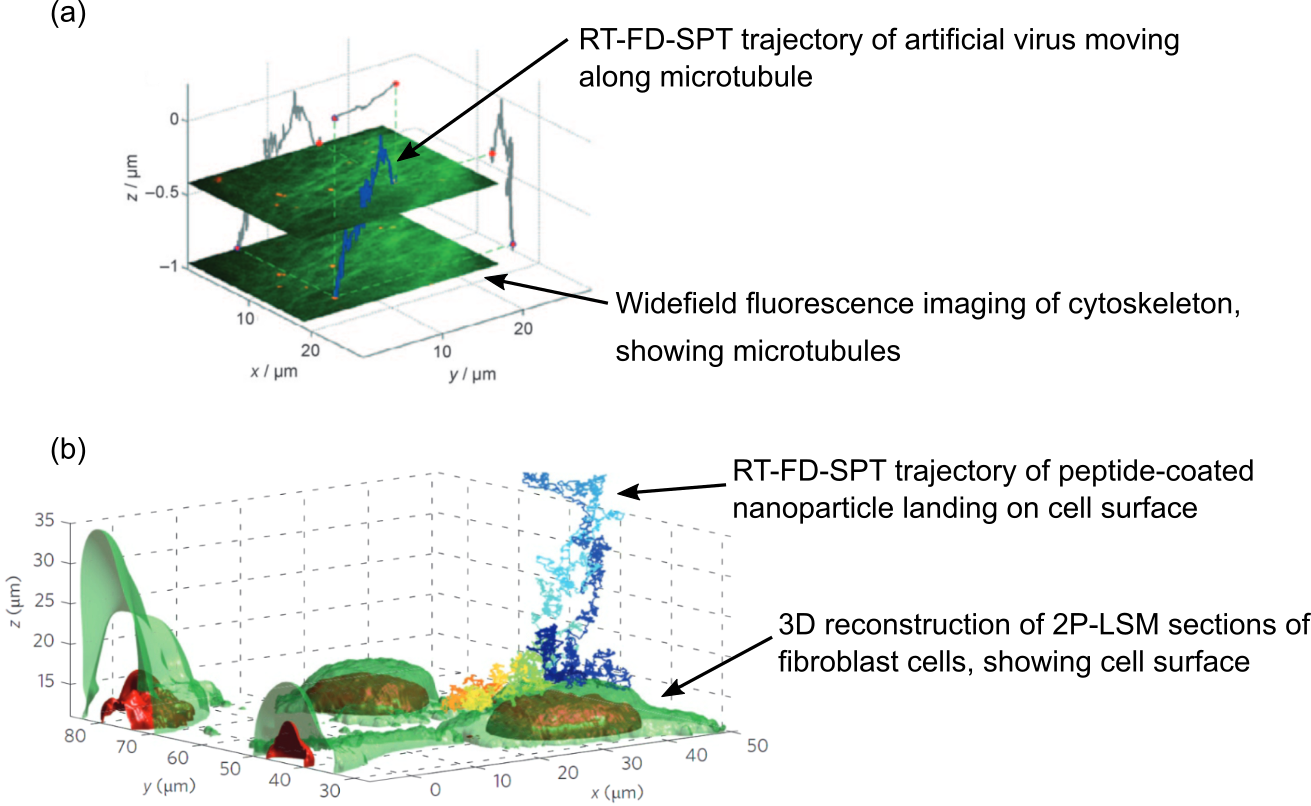}
        \phantomcaption
        \label{fig:microscopy_katayama}
    \end{subfigure}
    \begin{subfigure}[t]{0\textwidth}
        \includegraphics[width=\textwidth]{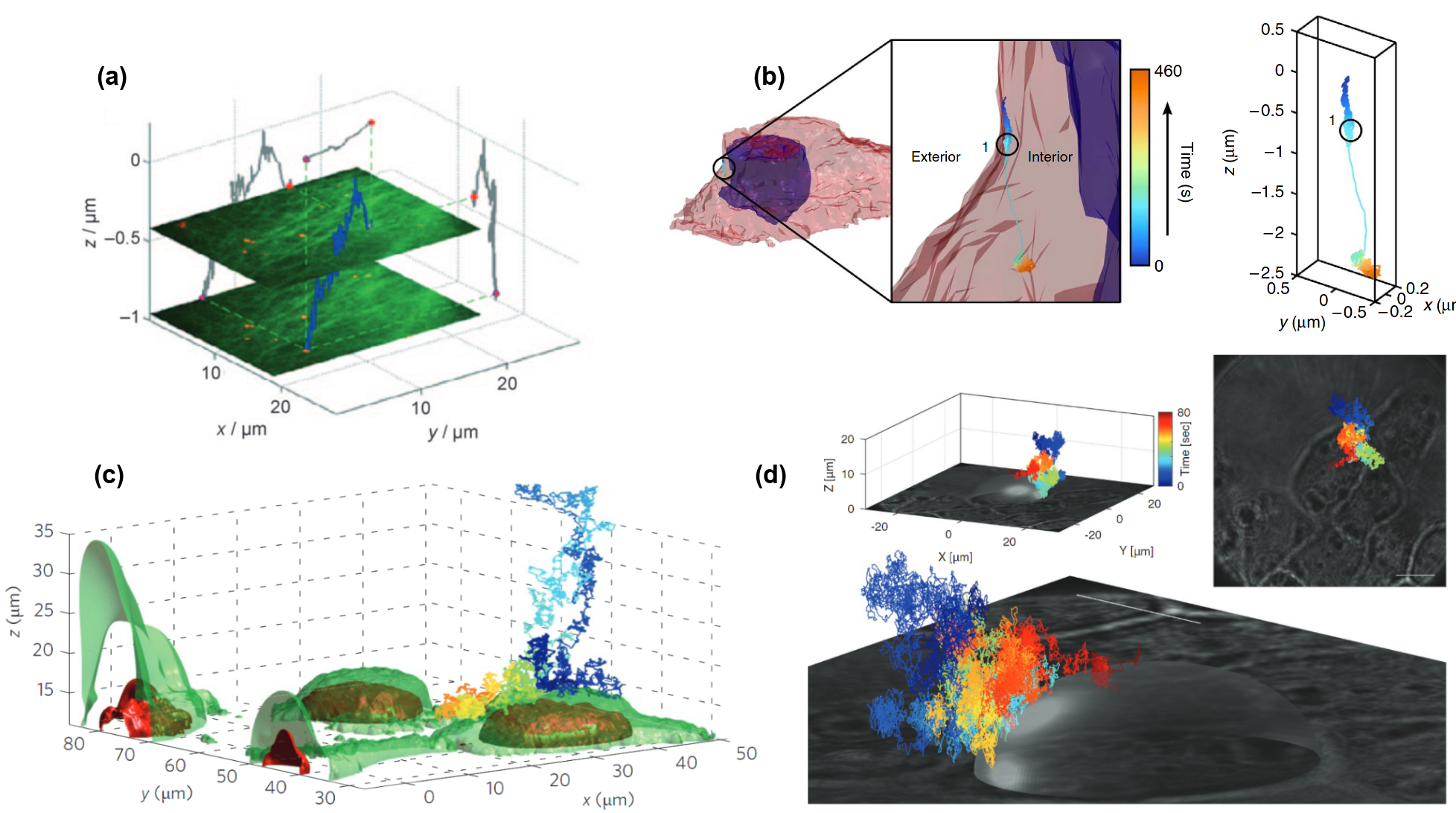}
        \phantomcaption
        \label{fig:microscopy_mrm}
    \end{subfigure}
    \caption{Example microscopy applications of RT-FD-SPT with concurrent contextual imaging. \textbf{(a)} Trajectory of an artificial virus in a live HUH7 human hepatome cell transfected with eGFP-tagged tubulin along with wide-field images at two different $z$-positions. RT-FD-SPT precisely measures the virus's trajectory, while the concurrent imaging shows the microtubules. This combination of techniques allows the experimenter to distinguish between movement along the microtubule and movement of the microtubule itself. Reprinted from Ref.~\cite{Katayama2009}. Copyright 2009 Wiley-VCH. \textbf{(b)} Trajectory of a PS-QD-Tat nanoparticle landing on an NIH-3T3 cell, with a 3D cell model from two-photon laser scanning microscopy (2P-LSM). Here, the 2P-LSM image shows the cell surface while the RT-FD-SPT trajectory shows the motion of the particle along that surface. Reprinted with permission from Ref. ~\cite{Welsher2014}. Copyright 2014 Macmillan Publishers Ltd.\label{fig:microscopy}}  
\end{figure}

\textbf{Figure~\ref{fig:microscopy}} shows two examples of RT-FD-SPT combined with concurrent imaging. In Figure~\ref{fig:microscopy_katayama} an orbital tracking trajectory of an artificial virus in a live cell is shown with concurrent widefield imaging at two different image planes~\cite{Katayama2009}. In this experiment, RT-FD-SPT allowed high-precision tracking of the virus over a 100 \textmu m range, while the concurrent imaging was used to distinguish between motion of the virus along a microtubule and movement of the microtubule itself. Figure~\ref{fig:microscopy_mrm} shows a multi-detector tracking trajectory of a peptide-coated nanoparticle as it landed on an NIH-3T3 cell, with the cells imaged using 2P-LSM. Here, RT-FD-SPT provided precision tracking at $\mu$s time resolution of a fast moving (2--5 $\mu$m$^2$s$^{-1}$) particle, while the 2P-LSM provided 3D visualization of the cell surface.

The main benefits of RT-FD-SPT over traditional SPT in all of these applications is a very long tracking range and good 3D tracking ability, with excellent spatiotemporal resolution. Some of the studies also exploited the ability to concurrently measure fluorescence spectra~\cite{Hellriegel2009} or intensities~\cite{Donovan2019}. The former proved useful as a way to distinguish the label of interest from a different fluorescent object. The latter was used in various ways to investigate the dynamics and interactions of the tracked particles. The advantage of the combination of RT-FD-SPT with other imaging modalities is illustrated by Welsher and Yang~\cite{Welsher2014} as a ``folding" of spatiotemporal regimes: RT-FD-SPT is able to capture dynamics at nanometer length scales and microsecond time scales, while concurrent imaging captures dynamics on micrometer length scales and second time scales.

\subsection{Spectroscopy-based applications}\label{sec:Spectroscopy}

SMS has proven to be a powerful tool for studying biological systems, allowing the detection of unsynchronized dynamics of molecular-scale systems that are averaged out in bulk measurements. It has been applied to the study of a wide variety of biological systems, including DNA~\cite{Ha2004,Alhadid2017,Kaur2019}, enzymes~\cite{English2006,Jiang2011}, molecular motors~\cite{Yildiz2004,Dienerowitz2021}, and photosynthetic proteins~\cite{Kruger2016,Kondo2017,Gruber2018}. With SMS, it is possible to observe, for example, conformational changes in proteins~\cite{Mazal2019}, transcription in single DNA molecules~\cite{Kapanidis2006,Hou2020}, enzyme reactions~\cite{Ha1999,Jiang2011}, and switches between various functional states in light harvesting complexes~\cite{Kruger2012, Schlau-Cohen2015,Gwizdala2016}. SMS is commonly used as an umbrella term for describing any type of single-particle detection (except imaging) using optical radiation, mostly fluorescence~\cite{Moerner2002}. Measurable time-dependent information-bearing parameters include the fluorescence brightness, spectrum, lifetime, polarization state, and photon correlations. First-order photon correlations enable different variants of FCS, while second-order correlations enable measurements of photon bunching and antibunching. Two-color fluorescence measurements allow the investigation of FRET. More advanced spectroscopic techniques such as excitation-emission spectroscopy~\cite{Thyrhaug2019} and ultrafast spectroscopy~\cite{VanDijk2005,Hildner2013,Maly2016,Liebel2018,Moya2021,Moya2022} have also been performed at the single-molecule level, as has Raman spectroscopy~\cite{Haran2010}. %(\textbf{Figure~\ref{fig:spectroscopy_techniques}}).
 
Some SMS methods have already been combined with RT-FD-SPT. Mabuchi's group introduced ``tracking-FCS"~\cite{Berglund2005} and used it to investigate DNA conformational fluctuations~\cite{McHale2009} . The same group also measured photon antibunching on diffusing quantum dots (QDs)~\cite{McHale2007}. Werner's group performed similar measurements on QDs~\cite{Wells2008,DeVore2015} %(\textbf{Figure~\ref{fig:spectrosopy_wells}}) 
as well as on fluorescent proteins~\cite{Han2012}. Yeh's group showed antibunching of dye-labeled DNA in solution~\cite{Liu2017} and in live cells~\cite{Chen2019} and Welsher's group showed antibunching of diffusing single dye molecules~\cite{Hou2020}. The latter group also used lock-in filtration of fluorescence brightness measurements to investigate the nanoparticle protein corona~\cite{Tan2021}. Werner and coworkers demonstrated the first direct measurements of fluorescence lifetimes with RT-FD-SPT~\cite{Wells2008}, and this capability was also shown by the group of Yeh~\cite{Perillo2015}, who applied it to the measurement of DNA hybridization kinetics in solution~\cite{Liu2017} %(\textbf{Figure~\ref{fig:spectrosopy_liu}}) 
as well as in live cells~\cite{Chen2019}. Similar measurements have also been performed on DNA origami structures using p-MINFLUX~\cite{Masullo2021}. 
Recently, Yang's group used lifetime information to improve their system's signal-to-background ratio by separating long-lifetime fluorescence from the short-lifetime background signal~\cite{Zhao2021}. Keller \textit{et al.}~\cite{Keller2018} used a total of eight detectors to perform single-molecule FRET (smFRET) measurements with simultaneous tracking on dye-labelled DNA. To our knowledge, the only measurements of emission spectra using RT-FD-SPT were done by Hellriegel and Gratton~\cite{Hellriegel2009}, who measured two-photon emission spectra of GFP-labeled uPAR in live cells . %(\textbf{Figure~\ref{fig:spectrosopy_hellriegel}}). 

Incorporating spectroscopic measurements into RT-FD-SPT typically involves additional detector hardware or an additional beam. To measure lifetimes, a TCSPC system synchronized with a fast pulsed laser needs to be added. The pulsed laser can also be used for tracking. For antibunching measurements, an additional point detector is needed together with correlation hardware, typically a TCSPC module. FRET measurements also require an additional detector. To measure spectra, a spectrometer is needed, which consists of a dispersion device (a diffraction grating or prism) and a line detector or 2D sensor such as a sensitive scientific CMOS or CCD camera. 

Yang and coworkers introduced the idea of a separate beam for spectroscopy ~\cite{Cang2006}, a technique that was subsequently also used by McHale and Mabuchi~\cite{McHale2009,McHale2010}. One advantage of this approach is that the tracking beam can have a different wavelength such as NIR, which can be used at a higher intensity without causing photodamage to the sample~\cite{Cang2006}. This could involve scattering or interferometric scattering (iSCAT)-based detection (see \textbf{Section~\ref{sec:curr_fut}}) or a separate tracking label to ensure that tracking and spectroscopy have separate photon budgets. For example, if one is interested in measuring spectra, all the photons in the spectroscopy channel could be sent to a spectrometer. For naturally fluorescent samples, a bright fluorescent label or the scattering signal could be used for tracking, while the comparatively weak natural fluorescence is investigated using spectroscopy. 
Another interesting use of two beams was the case of tracking-FCS measurements on DNA, where the tracking beam was larger than the molecule, whereas the probe beam was smaller than the molecule, making it sensitive to intramolecular motion, the study of which was the purpose of the experiments~\cite{McHale2009}. 

\begin{figure}
    \centering
    \begin{subfigure}[t]{0.8\textwidth}
        \includegraphics[width=\textwidth]{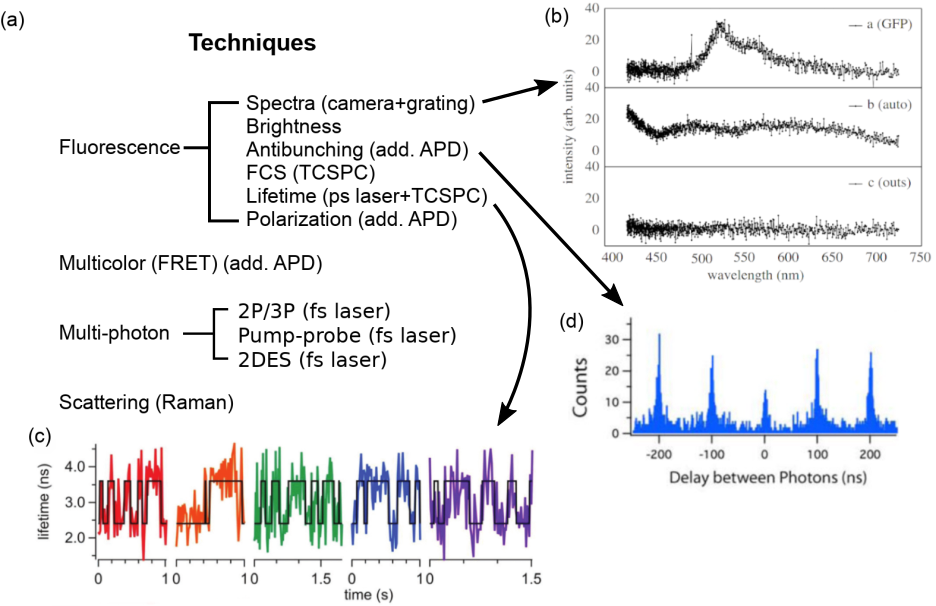}
        \phantomcaption
        \label{fig:spectroscopy_techniques}
    \end{subfigure}
    \begin{subfigure}[t]{0\textwidth}
        \includegraphics[width=\textwidth]{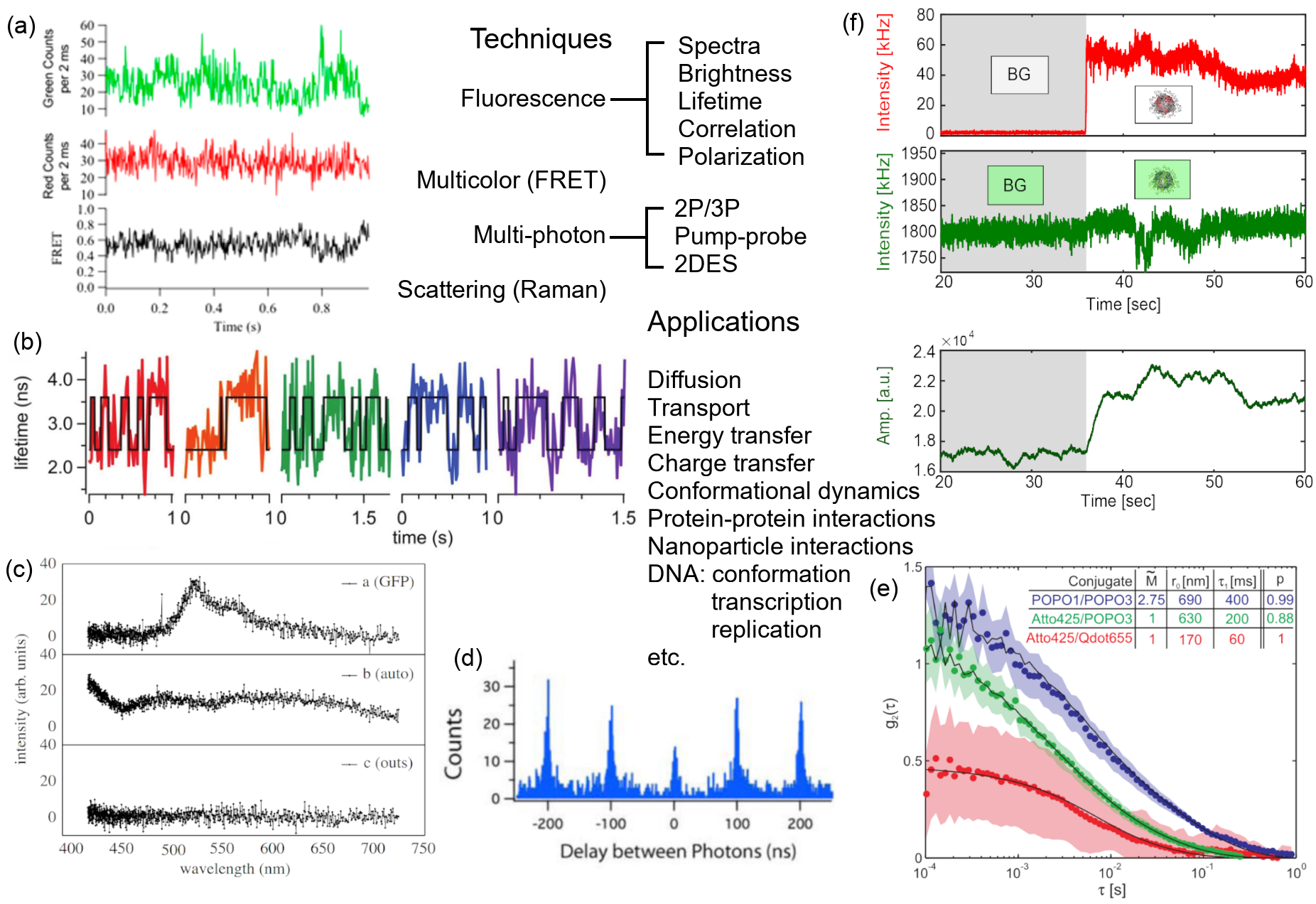}
        \phantomcaption
        \label{fig:spectrosopy_hellriegel}
    \end{subfigure}
    \begin{subfigure}[t]{0\textwidth}
        \includegraphics[width=\textwidth]{Figures/Spectro_overview.png}
        \phantomcaption
        \label{fig:spectrosopy_liu}
    \end{subfigure}
    \begin{subfigure}[t]{0\textwidth}
        \includegraphics[width=\textwidth]{Figures/Spectro_overview.png}
        \phantomcaption
        \label{fig:spectrosopy_wells}
    \end{subfigure}
    \caption{\textbf{(a)} Example spectroscopic techniques that could be used with RT-FD-SPT, with required additional hardware in parentheses. APD = avalance photodiode, FCS = fluorescence correlation spectroscopy, TCSPC = time-correlated single photon counting, FRET = Förster resonance energy transfer, 2P/3P = 2-photon/3-photon techniques, 2DES = two-dimensional electronic spectroscopy.  \textbf{(b)} 2-photon emission spectra from GFP-labeled uPAR in HEK293 cells. Adapted with permission from Ref. \cite{Hellriegel2009}. Copyright 2009 The Royal Society. \textbf{(c)} Fluorescence lifetime trace of labeled ssDNA, with switching indicating transient annealing/melting events. Reprinted with permission from Ref. \cite{Liu2017}. Copyright 2017 Royal Society of Chemistry. \textbf{(d)} Antibunching measurement of QD-labeled igE-Fc$\epsilon$RI on an unstimulated mast cell. Reprinted with permission from Ref. \cite{Wells2010}. Copyright 2010 American Chemical Society.}
    %Example applications of RT-FD-SPT spectroscopy. \textbf{(c)} \color{blue} 2-photon emission spectra from GFP-labeled uPAR in HEK293 cells. Adapted with permission from Ref. \cite{Hellriegel2009}. Copyright 2009 The Royal Society. \textbf{(d)} Intensity trace of tracked fluorescent nanoparticles with a protein corona. The red channel is the fluorescence from the NP while the green channel shows the fluorescence from the Alexa Fluor 488-labeled proteins. The grey region indicates the period when no tracking occurs. The bottom panel shows the 10 kHz lock-in filtration of the green signal, which removes the background to clearly show the presence of the protein corona. Reprinted with permission from Ref. \cite{Tan2021}. Copyright 2021 Wiley-VCH.  \textbf{(e)} Single-molecule FRET measurement of DNA labeled with Alexa Fluor 488 (donor) and Alexa Fluor 594 (acceptor). Reprinted from Ref. \cite{Keller2018}. Copyright 2018 American Chemical Society.  \textbf{(f)} Antibunching measurement of QD-labeled igE-Fc$\epsilon$RI on an unstimulated mast cell. Reprinted with permission from Ref. \cite{Wells2010}. Copyright 2010 American Chemical Society.  \textbf{(g)} Fluorescence lifetime trace of labeled ssDNA, with switching indicating transient annealing/melting events. Reprinted with permission from Ref. \cite{Liu2017}. Copyright 2017 Royal Society of Chemistry. \color{black} \textbf{(h)} Tracking-FCS measurement of genomic $\lambda$-phage DNA. Reprinted with permission from Ref. \cite{McHale2010}. Copyright 2010 Biophysical Society.}  
    \label{fig:spectroscopy_overview}
\end{figure}

\textbf{Figure~\ref{fig:spectroscopy_overview}} summarizes spectroscopic applications of RT-FD-SPT and shows a few illustrative examples of spectroscopic measurements that have been made with RT-FD-SPT. Figure~\ref{fig:spectroscopy_techniques} also indicates the additional hardware that is needed to add the technique to an RT-FD-SPT setup.   

Although the results so far are impressive, they have only scratched the surface of the wealth of information that could be obtained when combining RT-FD-SPT with SMS. To give just one, illustrative example of the impact that SMS studies can have on our understanding of the properties and functions of proteins and complexes at the nanometer scale, consider light-harvesting complexes. Owing to their autofluorescence, no fluorescent label is needed, enabling the use of fluorescence as a direct probe of the properties and protein conformational dynamics of the complexes. The light-harvesting complexes of a wide variety of photosynthetic organisms have been investigated, including bacteria~\cite{Bopp1999,Schlau-Cohen2013,Saga2002}--- especially cyanobacteria~\cite{Ying1998,Wang2015,Gwizdala2016,Wahadoszamen2020}---, plants~\cite{Kruger2010,Kruger,Schlau-Cohen2015}, and algae~\cite{Kruger2017}. Multiple spectroscopic parameters are typically used, often simultaneously. For example, correlations between changes in fluorescence brightness, lifetime, and spectra have been used to identify and characterize new physiological states~\cite{Kruger2010,Kruger,Schlau-Cohen2015,Gwizdala2016,Kruger2017,Gwizdala2018a}. It has also been demonstrated how the interactions between different individual proteins can be controlled~\cite{Gwizdala2018}, how metallic nanoparticles can drastically tune the photophysical parameters of these complexes ~\cite{Wientjes2014,Kyeyune2019}, and how ultrafast spectroscopy can resolve rapid fluctuations and quantum-coherent processes in these systems at the single-molecule level~\cite{Hildner2013,Maly2016}. RT-FD-SPT could be used to perform measurements such as these in a close to \textit{in vivo} environment or even in live cells. There would be drawbacks, the main ones being a decreased SBR and shorter measurement times due to tracked particles being lost. However, the advantage of the results being much more physiologically relevant would most likely outweigh these drawbacks for most experiments.

\section{Feedback Tracking and Control}\label{sec:Control}
%Section introduction and outline
The core of RT-FD-SPT is the tracking feedback control system. Feedback control enables the optical system to lock onto and track the target particle,  extending the duration over which one can take meaningful optical and spectroscopic measurements.  The design and implementation of the tracking controller as well as the tuning of its parameters can have a profound impact on the overall performance of the tracking system. In this section we discuss control strategies for RT-FD-SPT and discuss the challenges and limitations that must be considered when designing a controller for RT-FD-SPT.

\subsection{Controller Architecture}
\begin{figure}[htbp!]
    \centering
    \begin{subfigure}[t]{0.8\textwidth}
        \centering
        \caption{Single Detector Tracker}
        \includegraphics[width=\textwidth]{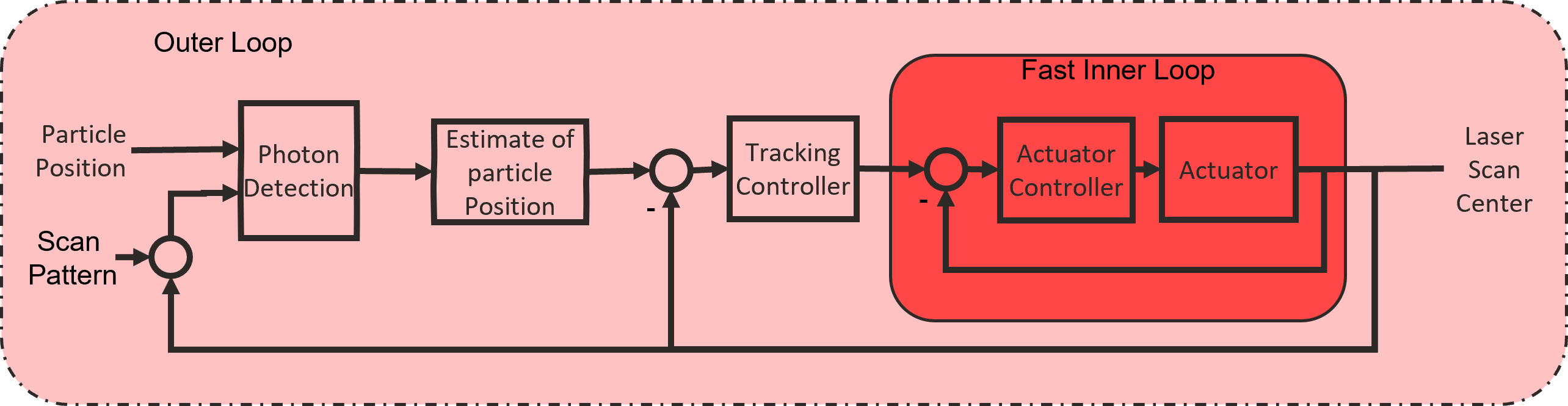}
        \label{fig:SingleDetectorTracker}
    \end{subfigure}\vspace{8pt}\\
    \begin{subfigure}[t]{0.8\textwidth}
        \centering
        \caption{Multi-Detector Tracker}
        \includegraphics[width=\textwidth]{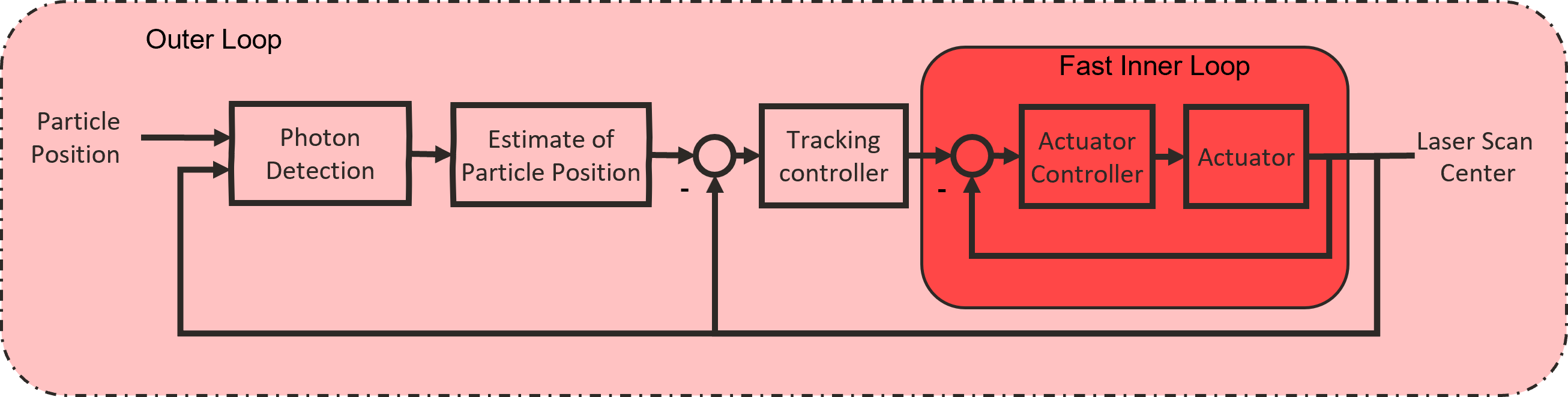}
        \label{fig:MultiDetectorTracker}
    \end{subfigure}\vspace{8pt}\\
    \begin{subfigure}[t]{0.8\textwidth}
        \centering
        \caption{Signal Maximization Tracker}
        \includegraphics[width=\textwidth]{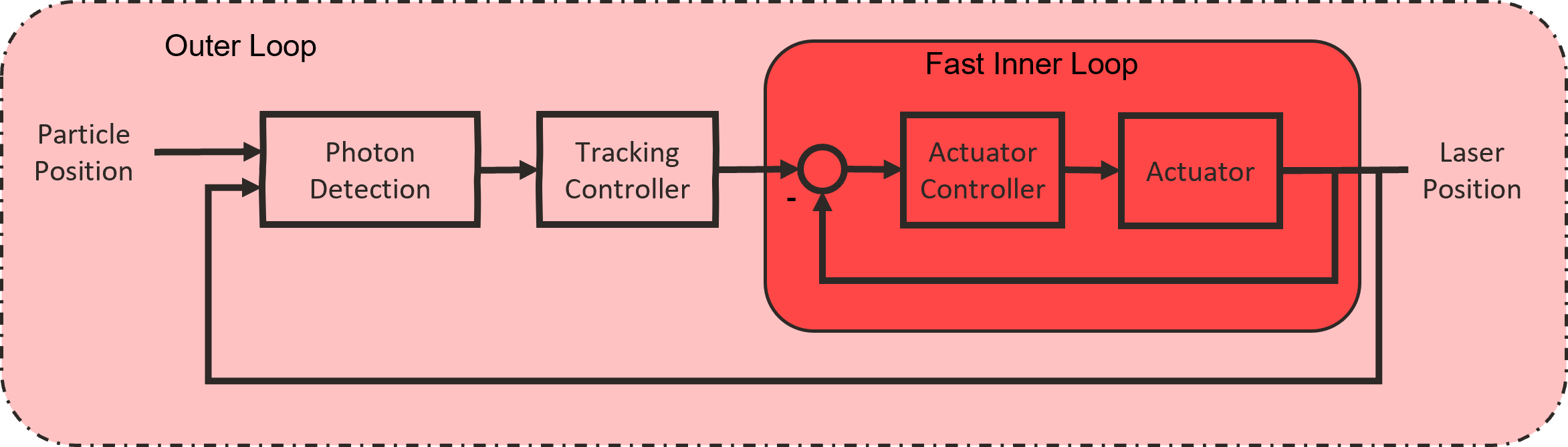}
        \label{fig:ESCTracker}
    \end{subfigure}
    \caption{System block diagrams for typical reference point RT-FD-SPT. Both the (a) single and (b) multiple detector systems feedback an estimate of the particle location while the (c) maximization-based approaches actuate based purely on the measured signal.}
    \label{fig:ControlSystems}
\end{figure}

There are two main architectures for control in RT-FD-SPT, one based on reference point tracking (\textbf{Figures~\ref{fig:SingleDetectorTracker}-\ref{fig:MultiDetectorTracker}}) and one based on a signal maximization approach (\textbf{Figure~\ref{fig:ESCTracker}}). Both rely on feedback to steer the tracking system and thus inherit the benefits that feedback brings, including robustness to noise, to external disturbances, and to changes in the system dynamics. In both of these architectures, at the innermost layer is the actuator hardware (which could be, for example, a piezoelectric stage, EOD, AOD, or a TAG lens) to realize the fastest time scale motion. These actuators typically come with manufacturer-designed controllers and for the purposes of this discussion are assumed to be a black box with an output that faithfully follows its input.

In the reference point tracking approach, the reference point is given by the estimated position of the particle (see \textbf{Section~\ref{sec:estimation}} for details on generating such estimates from the optical measurements). This scheme has been used in multi-detector tracking methods, where the actuated variable is typically the center of the detector array, and in single detector techniques, where the actuated variable is the center of a scan pattern. The goal of the controller is to drive the actuated variable to the reference point and the design of this controller is (at least conceptually) independent of whether the system is moving the sample towards a fixed excitation/detection setup or moving the excitation/detection setup towards the sample. 

A common controller choice in the regulation approach is a Proportional-Integral-Derivative (PID) controller. PID controllers have been implemented with good results in local imaging tracking \cite{Juette2010}, local scanning tracking \cite{So1997SPT, Ragan2006, Hou2017, Hou2019,Hou2020}, multi-detector tracking \cite{Cang2006, Cang2007}, measurement constellations \cite{Balzarotti2017, Perillo2015, Liu2015, Liu2020a}, and orbital scanning tracking \cite{Gratton2004, Gratton2005}. One of the major reasons for the success of PID is that there is a simple interpretation for each of the three components of the controller, allowing intuitive adjustment of the parameters to achieve good performance. The P term applies a feedback signal that is directly proportional to the measured error and, in general, increasing this gain leads to a faster response with lower final error. However, making it too large can drive the system to ``overshoot'' the goal position and create a strong oscillatory response with a long transient. The I gain serves to reduce the final error of a system but can lead to instability (where the system does not track the particle) if tuned too high. The D term provides an anticipatory response, reacting as soon as the error begins to change, with a magnitude that is proportional to how fast it is changing. However, derivative action tends to amplify any noise in the signal and can lead to undesirably sharp responses that can excite nonlinearities in the actuators. Because of these downsides, in practice many applications use either just P control \cite{Gratton2003,Gratton2004, Gratton2005}, PI control \cite{Balzarotti2017,Perillo2015,Liu2015, Liu2020a, Hou2020}, or even just I control \cite{Hou2017}. A deep and thorough discussion of PID control can be found in \cite{aastrom2006advanced}.

Achieving good tracking performance on a specific system depends on good tuning of the controller parameters. Many of the published algorithms in RT-FD-SPT involve hand-tuned parameters, which are adjusted until satisfactory performance is achieved based on either physical experiments or simulations \cite{Liu2015}. However, this approach can be quite time-consuming and the level of performance achieved depends strongly on the skill of the person doing the tuning. For PID control, an alternative is to apply one of a variety of heuristic methods that use easy-to-measure experimental values (such as a rise time or the period of an induced oscillation) to calculate gain values that provide a good balance of speed and stability for the particular system. Many such rules have been developed (see, e.g., \cite{PIDhandbook}) with perhaps the best known being that of Ziegler-Nichols (see, e.g., \cite{franklin2019feedback}). While some amount of hand-tuning is typically still needed after applying a heuristic, applying these rules ensures that this tuning starts from values that already provide reasonable performance.

While PID often works well in practice, it is a fairly conservative controller, providing stability and good robustness against noise, disturbances, and uncertainty at the cost of speed of response. There are several alternatives based on model-based control that can outperform PID but which, as the name implies, require an accurate model to ensure good performance and run the risk of causing instability if those models are not reasonably close to the true system. Developing controllers based on these techniques often requires a strong background in control engineering and thus these schemes are outside the scope of this review. One exception worth noting is that of the Linear Quadratic Regulator (LQR). LQR develops a controller designed to minimize a cost function given by a weighted combination of the tracking error (seeking to drive it to zero) and the magnitude of the control signal (seeking to keep it small). While more challenging to use than simple PID, LQR has the benefit of a built-in notion of optimality and the ability to easily modify the controller to ensure the control signal does not hit physical constraints such as voltage bounds or actuator slew rate limits (though without providing a theoretical guarantee on satisfying any such constraints). It has been applied to good effect in several RT-FD-SPT algorithms \cite{Berglund2004, Shen2009,Shen2009a, Shen2011, Shen2011a}.

The second class of controllers, based on signal maximization, are inherently nonlinear but come with the benefit of not relying on any estimate of the particle position. Rather, they work directly with the measured signal, steering the measurement volume either to maximize that signal or to converge to a periodic pattern with a constant signal level. An early version of this idea was used to automatically find and position a laser for spectroscopic studies of single molecules \cite{Ha1997SPT}.  The controller in this case used a simplex method to measure the photon count rate, and used a gradient ascent to align the laser to the molecule automatically. More recently, an extremum-seeking controller was developed to converge to an orbit around a fixed fluorescent particle \cite{Andersson2010, Andersson2011, Ashley2012, Ashley2016, Ashley2016a, Ashley2018, Pinto2021a}. Through simulations and experiments, it was shown that this controller naturally switches between two distinct behaviors, converging to the desired orbital motion for slow moving particles and transitioning to a chasing behavior as the speed of the particle increases. While the ability to operate directly on the intensity signal, thereby separating particle localization from control, is appealing, the nonlinear nature of these approaches makes theoretical analysis and parameter tuning more difficult than PID. Still, the benefits of these nonlinear approaches can be valuable. For further guidance on the convergence properties of the extremum-seeking approach and on selecting controller gains, see \cite{Ashley2018,Pinto2021}. Ultimately, the decision to select a signal maximization method, with its added complexity to tune, due to its nonlinear nature, but ease of implementation, due to the lack of an online estimator, depends on a variety of factors. Unfortunately, there is no existing detailed, experimental comparison between these methods and those based on reference point tracking; this represents an opportunity for future research.

One often overlooked aspect common to all control architectures is the initialization step, namely how to select a particle for tracking. There are two primary methods for doing this: image-based and threshold-based. Under the image-based approach, an image of the sample is captured and displayed to the user. That user then selects a particle of interest to begin the tracking process~\cite{Shen2009a}. This allows the user to ensure ``interesting'' particles are selected but can be severely constraining in terms of particle speed, since the particle cannot move significantly in the time it takes to perform the capture, display, and selection steps. In practice, this method works for particles with diffusion coefficients of about 0.1 $\mu$m$^2$ s$^{-1}$ or less (see Supporting Information for more details). Under the threshold-based approach, the user selects a location but not any individual particle. The detection volume is moved to that location and monitors the photons counts. When a sudden increase over a given threshold is detected, the tracking controller is enabled~\cite{Balzarotti2017}. Since selection is no longer constrained by user reaction time, faster particles can be tracked. In addition, once a particle is lost, the system can switch back to dwell mode and wait for the next threshold crossing, automatically picking up another particle. However, in this approach, background noise and detector counts force a lower bound on the threshold to avoid too many false detections, possibly making the system somewhat insensitive to dim particles. While one can apply a systematic approach such as constructing a likelihood ratio test to determine a threshold that balances sensitivity against false detections, in practice the value tends to be chosen through trial and error. In addition, many fluorophores exhibit blinking or temporary dark states.  Enabling the system to wait for a short time for the particle to enter a bright state can extend the duration of tracking a particle.

\subsection{Control Challenges and Limitations}

\begin{figure}
    \centering
    \includegraphics{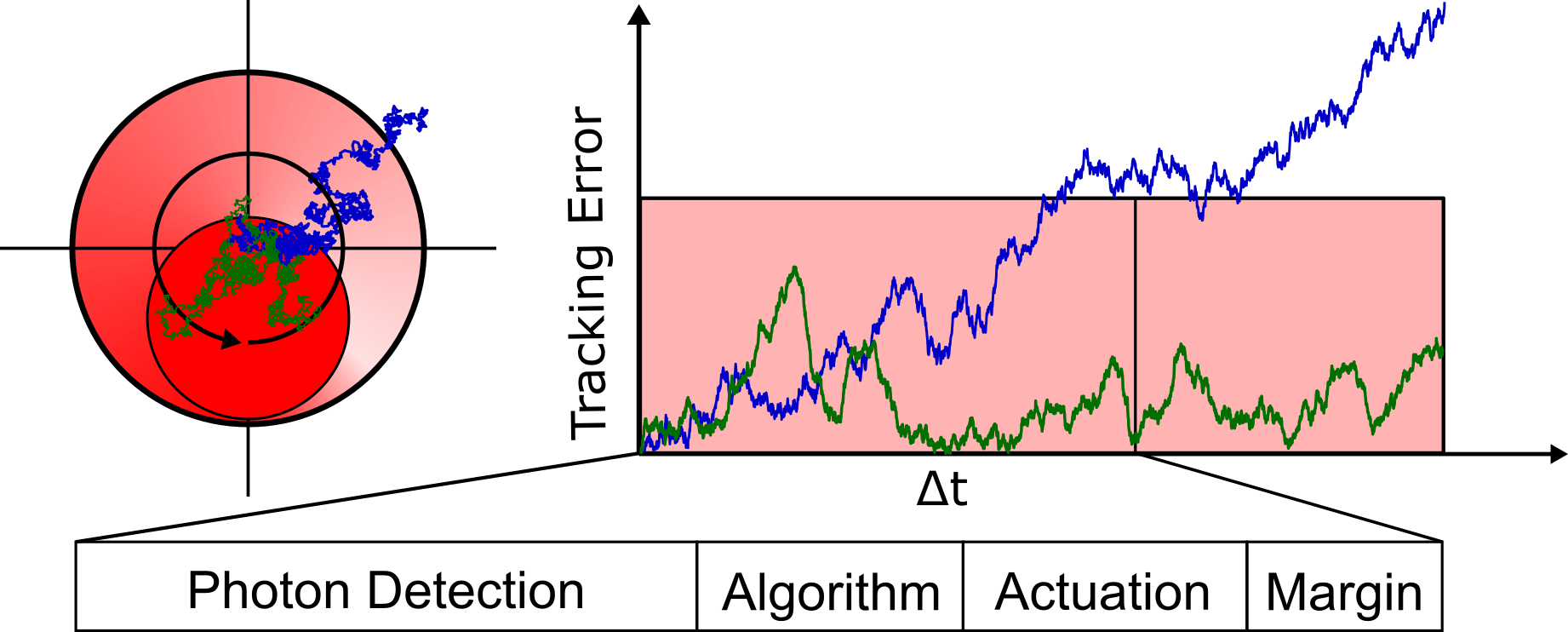}
    \caption{Temporal budget. Inside each update step of duration $\Delta t$, any RT-FD-SPT algorithm must collect photons from the tracked particle, translate these photons to a desired update of the tracker, and then implement this update through actuation. Under stable tracking, the algorithm will keep the particle inside the detection volume of the algorithm (green trajectory); if the particle is too fast relative to the update rate, or the estimation error too large, then the particle may move out of the detection volume before an update is completed, leading to loss of tracking (blue trajectory).}
    \label{fig:TemporalBudget}
\end{figure}

The performance of RT-FD-SPT algorithms is ultimately limited by several physical challenges and trade-offs that must be made when designing the control system. To help understand these trade-offs, we organize this discussion around a time budget as shown in \textbf{Figure~\ref{fig:TemporalBudget}}. Within each update step of the controller there are three steps that must occur: detection and collection of the photon signal, implementation of the algorithm to determine the actuator command, and then, finally, the actual motion of the physical hardware in response to that command. It is also important from a practical implementation point of view to include a margin in the update time where no work is done; this allows the system to absorb small jitters in the timing and allow the computational hardware time to perform other functions (such as storing data or communicating to other systems in the experimental setup). The update rate must therefore be slow enough to allow enough time to perform all three steps, as well as to acquire any spectroscopic data desired, while also being fast enough to track the particles of interest over a long enough duration to acquire sufficient measurements to probe the questions of interest. Note that most, if not all, RT-FD-SPT approaches make the assumption that the particle is quasi-static, that is, that the time scale of the algorithm (as defined by this update rate) is much faster than the motion of the particle.

Satisfying this quasi-static assumption is important to achieving the expected performance of RT-FD-SPT. An upper bound on the maximum speed of particle dynamics that can be reliably tracked can be developed by assuming that the particle must stay within the detection volume within a single time step. If the particle moves out of this volume prior to the next measurement, then that measurement will yield no information about the particle and it is lost. The maximum distance traveled within one time step is described by a probability distribution defined by the particular dynamics of the particle. Assuming the particle starts at the center of the detection volume at each step (perfect tracking), the notion of the first passage time to reach the edge of the detection volume sets an upper bound on the particle speeds that can be tracked \cite{Andersson2005,Ashley2018}. This upper bound can be increased either by increasing the update rate or by increasing the detection volume. The update rate is limited by the time required for the three steps of RT-FD-SPT; the limitations for each of these steps are described further below, with typical values for completing all three steps ranging from 125 $\mu$s \cite{Balzarotti2017} to 5 ms \cite{Wells2010,Perillo2015}. Increasing the effective detection volume can be done by expanding the size of the scan or constellation used in the different approaches, with typical length values ranging from about 400 nm \cite{Balzarotti2017} to about 3 $\mu$m \cite{Hou2017}.  For example, the work in \cite{McHale2007} expanded the size of their orbital scan to enable antibunching measurements on freely diffusing fluorophores with diffusion coefficients of 20 $\mu$m$^2$ s$^{-1}$; this came at the cost, however, of a loss in localization performance with an accuracy on the order of 200--300 nm. This trade-off exists on a continuum and it can be shown (see Supporting Information) that the loss in localization performance is proportional to the square of the effective size of the detection volume when keeping detected photon counts constant and to the fourth power of that size when keeping the illumination power constant. It is also important to note that collecting photons over a larger effective detection volume implies a larger distance to be covered for scanning approaches, necessitating either very fast scanners (as in, e.g., 3D-DyPLoT and its variants \cite{Hou2017,Hou2019,Hou2020}) or a larger portion of the time budget allocated to photon collection. Of course, meeting a desired speed of tracking is only one element in selecting the update rate and how to spend the time budget it defines. It is important to understand the trade-offs involved in giving more or less of that time budget to each of the steps in the update.
 
\underline{Photon collection:} The first step in an update is the collection of photons. Enough informative photons must be collected for either the online localization algorithm to provide accurate estimates of the particle location or for the maximization-based controllers to accurately estimate the direction of signal increase. Because photon detection is fundamentally a Poisson-distributed counting process \cite{ram2006stochastic, Gallatin2012, chao2016fisher, mertz2019introduction}, the acquired signal is always noisy with a Signal-to-Noise Ratio (SNR) that improves as $\sqrt{N}$, where $N$ is the number of photons detected. The SNR can be improved by increasing the (average) number of photons acquired, either by increasing the photon emission rate or by increasing the time spent collecting photons. The first approach is limited largely by concerns of photobleaching and phototoxicity and is often not a viable approach when using many fluorescent dyes or autofluorescent proteins. The second approach comes at the cost of reducing the time available to the other steps in the update or in decreasing the update rate and slowing down the overall controller. The effectiveness of this is also reduced by the reality that the particle is in constant motion during the photon acquisition process, leading to motion blur in the data that degrades overall performance. While some estimation algorithms include at least partial efforts to overcome motion blur \cite{Berglund:2010ffa,calderon2016motion}, it remains a challenge in RT-FD-SPT and is one of the drivers for making the quasi-static assumption. 

In most RT-FD-SPT approaches, measurements from multiple locations are needed at each time step, further dividing the time budget, as all of this data must be collected inside a single update. Most implementations, then, place at least some emphasis on how fast data can be acquired from different locations, some by using multiplexed detection \cite{Cang2006, Cang2007, Cang2008, Juette2008, Juette2010, Juette2013} or pulsed interleaving \cite{Perillo2015, Liu2017, Liu2020}, others by employing high-speed AOD \cite{McHale2007} or EOD scanners \cite{Balzarotti2017, Hou2017}, and still others by using advanced control methods to achieve fast motion \cite{Shen2011a,Andersson2011}. Note that when using these methods, multiple cycles of photon collection can and does occur during the time budget. This provides a high temporal resolution (also known as time to localization) down to 20 $\mu s$. Unfortunately, this temporal resolution is often conflated with the update rate of the tracking controller, which often occurs at a slower rate ($\sim\! 5$ ms) due to the smaller bandwidths of the actuators used for tracking \cite{Cang2006, Perillo2015, Hou2017, Balzarotti2017}.

More recently, some approaches recognize that not all photons are equally informative with respect to localizing the particle. Using ideas from optimal estimation theory, particularly the Fisher Information Matrix and the corresponding Cramér-Rao lower bound, one can ensure the photons that are collected are useful with respect to the localization task \cite{Gallatin2012} (see Section~\ref{sec:estimation}). This idea has been shown to improve precision by approximately 20\% in 3D localization relative to uniform sampling \cite{Zhang2021} and used to design control algorithms for steering the excitation source \cite{Vickers2021}.

Finally, one must also consider how photon collection affects other aspects of estimation and system performance. The detection pattern used to localize the molecule must provide data that enables the localization algorithm to produce a unique result. In most methods, a symmetric pattern provides this. However, the shape of the beam used in MINFLUX requires an additional measurement at the center of their circularly-symmetric pattern to ensure a unique solution \cite{Balzarotti2017}.  Another aspect to consider is the localization uncertainty compared to the characteristic width of the detection volume. In essence, the total uncertainty from all sources, including localization error, motion blur, and particle dynamics, must be smaller than the width of the detection volume to ensure tracking is maintained. Additionally, while the focus is generally on tracking, the information obtained is often used to estimate other parameters. While image-based approaches to SPT have addressed the effect of localization uncertainty on such estimation through both Fisher information \cite{vahid2020fisher} and Monte Carlo simulations \cite{ober2022limits}, this remains an open question in RT-FD-SPT.

\underline{Algorithm:} The second step in an update is to perform all the calculations necessary to translate the measurements into an action. For most schemes, this includes filtering to reduce the effect of noise in the signal, online estimation of the particle position, and then calculation of the control signal. Note that some level of filtering is important since, as noted above, the photon signal is inherently noisy. The fluctuations in these measurements persist through the estimation and thus are injected into the feedback loop of the controller. It was shown experimentally in Ref. \cite{Hou2019} that a tracking controller that is more sensitive to changes in the reference point shows an increased localization uncertainty for a static particle. When the controller is made less sensitive, the static particle localization precision improves, but at the cost of speed of response. Filtering of the signal using, e.g., a Kalman filter \cite{Berglund2004, Andersson2005, Shen2009, Hou2017}, helps to mitigate the noise, potentially allowing for higher gain controllers, but needs additional computation in the update step. Note that the maximization-based approaches avoid the need for online estimation but may still benefit from filtering. In order to keep the computation time as short as possible, many techniques use linearized estimators \cite{Berglund2004, Balzarotti2017} or simple difference-over-sum estimators \cite{Enderlein2000a, Wells2008, Perillo2015} to reduce the computational load, combined with fast hardware such as lock-in amplifiers \cite{Berglund2005} or field-programmable gate arrays (FPGAs) \cite{Balzarotti2017, Hou2017} to do the calculations as fast as possible.

\underline{Actuation:} The final step in an update is to apply the control signal to the actuation hardware and for the system to complete the desired motion for the current update. The time scale for this step is defined by the bandwidth of the actuation system. In many cases, particularly with piezoelectric components and their manufacturer provided controllers (which generally value robustness and safety over speed), this is the bottleneck limiting the overall speed~\cite{Welsher2015}. Significant improvements in bandwidth can be achieved by designing custom low-level controllers but doing so requires both access to the direct inputs to the actuators and experience in control systems design. In general, then, selecting fast hardware and making the most out of that hardware through appropriate tuning of the low-level controllers and of the algorithm parameters helps to minimize the impact of this step on the overall time budget. One major challenge here, however, is that the gains that give the best tracking performance are a function of the experimental conditions, including considerations such as the speed of the particle, the photon detection rate, and the background count rate.
For example, it was shown through simulations that the tracking error of TSUNAMI depends on both the controller gains and the diffusion coefficient of the tracked particle \cite{Liu2015}. The best gains are a function of diffusion coefficient; gains that were too large amplified the estimator noise while gains that were too small yielded insufficient sensitivity to changes in the particle location. Even more challenging, the experimental conditions typically change during the course of experiment. Two common examples of this are photobleaching, causing a reduction in the detected photon rate over time, and changes in the diffusion coefficient due to binding events or changes in molecular crowding \cite{Welsher2014}. In these cases, the best tuned controller can become ill-suited to the changing experimental conditions. 

To overcome these challenges, controller parameters can be adapted in real time in response to changing conditions. In some approaches, the size and shape of the scan pattern  is modified based on the current uncertainty about the particle location; in essence, widening the scan pattern improves detection while tightening the scan pattern  improves localization precision \cite{Balzarotti2017, Gwosch2020}. Note that for beam scanning methods, adapting the scan pattern also requires adapting the beam diameter to maintain good localization \cite{Gallatin2012, Vickers2021}. Other methods have adjusted the scanning volume based on the (estimated) diffusion coefficient of the particle, increasing localization precision when that coefficient is small and increasing the probability of detection when it is large \cite{Hou2019}. The decision to adapt based on localization uncertainty or based on diffusion coefficient depends largely on which source of error dominates a particular application. While the use of such adaptive controllers is still limited, they show promise for improving the overall performance by adjusting to experimental realities.

\section{Estimation and Fisher Information}\label{sec:estimation}

\subsection{Estimators} 
%introduction of section
Estimation is a key component in RT-FD-SPT. It plays two related but distinct roles: to provide the position of the particle in reference point tracking controllers (see \textbf{Section~\ref{sec:Control}}), and to provide information to fulfill the scientific aim. The specific scientific aim is, of course, application dependent and includes estimating the position of the particle and its diffusion coefficient \cite{Ashley2015, Ashley2016a, Hou2019}, detecting when the diffusion coefficient changes \cite{Welsher2014,Godoy2021}, determining dynamics of conformational changes \cite{Watkins2004}, and estimating changes in photophysical states \cite{Watkins2005}. 

Algorithms intended for direct use in the feedback controller use online estimation and are thus restricted to causal techniques as they can only use measurements up to the current time. It is also vital that the estimator complete its computations quickly, both to support a fast closed-loop rate in the tracker (see \textbf{Section~\ref{sec:Control}}) and because estimation algorithms almost invariably assume the particle is static during the measurement process. Many scanning-based or multidetector-based approaches also assume that the tracker keeps the particle close to the center of the pattern, allowing for a linearization of the observation model around this point so as to develop efficient estimators. Because speed is typically one of the primary considerations in the controller, it is often acceptable to sacrifice some accuracy for speed, relying on the robustness of the controller to handle the resulting error. At one end of the trade-off between accuracy and speed are simple centroid-style algorithms \cite{So1997SPT,Ragan2006, Juette2008, Juette2010, Juette2013} and difference-over-sum estimators \cite{Cang2006, Cang2007, Cang2008, Wells2008, Han2012, Keller2018, Welsher2014, Welsher2014, Welsher2015}. Slightly more complex are algorithms that take some advantage of knowledge about the shape of the point spread function (PSF) of the instrument, including the fluoroBancroft algorithm \cite{Shen2009, Shen2009a, Shen2011, Shen2011a, Shen2012} and assumed density filters \cite{Fields2011, Hou2017, Hou2019, Hou2020}. At the other end of this trade-off are methods based on optimal estimation, including maximum likelihood estimation (MLE) \cite{Andersson2005,Balzarotti2017,Vickers2021}, minimum mean squared error estimation (MMSE) \cite{Balzarotti2017}, and least squared error estimation (LSE) \cite{So1997SPT,Ragan2006, Andersson2007,Shen2009,Shen2009a,Shen2011,Shen2011a,Shen2012,Bewersdorf2011}. In many cases, the estimation algorithm is coupled with a filtering scheme (often a Kalman filter) to help remove noise from the estimates \cite{Andersson2005,Shen2009,Shen2009a,Shen2011,Shen2011a,Shen2012,Fields2012,Hou2017,Hou2019,Hou2019a,Hou2020}. 

While the estimation error has a clear impact on the quality of tracking, its effect on the feedback controller in a RT-FD-SPT system has not been explicitly considered in the literature. As noted in the previous section, a simple upper bound on the allowable error can be defined by ensuring that it is smaller than the radius of the detection volume of the system since otherwise the estimation error will quickly drive loss of tracking. While adding filtering can reduce the estimation error, this comes with its own cost through a slowing of the responsiveness of the controller. The right balance of these considerations is, of course, defined by the particular application and experimental aims.
%The amount of acceptable estimation noise has not been addressed in the literature. In practice, we do not want the particle to exit the detection footprint, or else we lose the particle. A guide to help initially set an upper-limit is to use the worst case error. This is the sum of the mean square tracking error, the maximum estimation error, and the distance traveled by the particle in a single time step. This sum must be less than the radius of the detection footprint so that the system can continue tracking the particle. Of course there are other factors to consider as well, including the inclusion of filtering schemes, the responsiveness of the controller, and the particular experimental aims. 

Online estimation has been shown to be a powerful tool for providing position estimates to the tracking controller, and a well-designed and well-tuned feedback controller can often overcome the errors introduced by limited-quality online estimation. 
These controllers act directly on the measurement, comparing it to the expected (or desired) value and acting to reduce that error. If the estimator is unbiased, then on average, the controller drives that error to zero. If the estimator is biased it is still possible to drive the error to zero if in each cycle of the controller that error is reduced~\cite{Ernst2012}.
%\color{blue}This is illustrated by the following.  From Ernst et al~\cite{Ernst2012}, the online estimator has bias where the estimated error is smaller than the true error.  The system moves towards the estimated position of the particle. This position is not at the true location of the particle but is closer than the previous position. The true error has been reduced on average. The principle of feedback repeats this process; each iteration brings the system closer to the particle until the error is dominated by the estimation uncertainty. \color{black}
In addition, this feedback provides additional robustness to the measurement noise, disturbances (both external and from estimation error), and modeling error (such as changes in the actuator dynamics across their workspace) that are always present in any physical system. As noted above, the need to make the estimator computationally cheap necessitates making simplifying assumptions. This can reduce performance as those assumptions do not hold exactly in practice. For example, the linearization of the observation model results in a mismatch between the actual optical model and the model used for estimation \cite{Ernst2012, Balzarotti2017}. This mismatch increases as the particle moves from the point of linearization, systematically increasing the bias of the online estimates~\cite{Ernst2012}. By contrast, offline estimation can apply algorithms of much higher computational complexity which are not limited to causal estimation, allowing the entire set of data to be used in making estimates at every point in time. To the best of our knowledge, only two groups currently use offline estimation. The first is the work in~\cite{Balzarotti2017}, where the online estimates used for tracking were numerically unbiased after tracking had concluded to improve the overall localization accuracy. The second is in the signal maximization approach \cite{Ashley2016a,Ashley2018}. Because this approach reacts directly to the intensity, there is no need for online estimation; in addition, the pattern-less trajectory is ill-suited to many of the estimation approaches noted above. To handle the particular stream of data produced by this approach, an algorithm that combines nonlinear filtering with an optimization-approach to jointly estimate particle location and model parameters was developed \cite{Ashley2015,Ashley2016a}.  

\subsection{Fisher Information and Cramér-Rao Lower Bounds}
Regardless of the estimation method employed, there is a fundamental limit to the achievable precision in parameter estimation. This minimum achievable estimator error variance for any unbiased estimator is known as the Cramér-Rao Lower Bound (CRLB)  (see, for example, \cite{casella2021statistical,zacks1971theory}). The CRLB is determined from the Fisher information matrix FIM, defined by
\begin{equation}
    \textrm{FIM} = \mathbb{E}\left[\left(\frac{\partial \ln(p(y \lvert \theta))}{\partial \theta}\right)^T \left(\frac{\partial \ln(p(y \lvert \theta))}{\partial \theta}\right) \right],
\end{equation}
where $\ln(p(y \lvert \theta))$ is the log-likelihood function of the recorded signal $y$ and $\theta$ is the vector of parameters one wants to infer. The CRLB is then given by the corresponding diagonal elements of the inverse of the FIM:
\begin{equation}
   \textrm{Var}(\theta_i) \geq [F^{-1}]_{ii}.% \leq [F]_{ii}^{-1}
\end{equation}
Understanding the CRLB for a particular experimental setting can help in gauging the quality of a particular experiment. Of course, as this is a theoretical limit, practical estimators typically perform worse than the CRLB so it should not be interpreted as the variance in a specific experiment but as a \textit{guide} as to what can be achieved. Note that the use of MLE is in part justified because such estimators achieve the CRLB in the limit of large amounts of data. 

While the CRLB directly gives the lower bound on the achievable precision, the FIM can be understood as representing the information content contained in the measurements. A ``larger'' FIM (in an appropriate sense) implies a larger amount of available information. This can translate into being able to meaningfully measure smaller changes in the parameter of interest. Similarly, increasing the number of parameters to be estimated will increase the CRLB for any one of the parameters; in essence there is a finite amount of information in the data and increasing the number of estimated parameters ``spreads'' that information over a larger set. As a result it is always better to directly measure as many parameters as possible to ensure the highest quality estimation in the parameters that matter.

The first comprehensive application of FIM in microscopy was by Ram \textit{et al.} \cite{ram2006stochastic} where the framework was applied in a consistent and detailed way to derive the CRLB for particle localization in widefield super-resolution microscopy (a comprehensive tutorial on Fisher information for optical microscopy published by the same research group \cite{chao2016fisher} is a good source for further information). The FIM has found utility as an analysis tool for comparing different widefield microscopy techniques \cite{zhou2019advances} and as a design tool \cite{shechtman2014optimal}. It has also been applied in the RT-FD-SPT setting, beginning with the work of Gallatin and Berglund \cite{Gallatin2012} where they showed that the FIM for particle localization based on laser scanning depends on the electric field gradient of the laser: increasing this gradient increases the FIM and thus decreases the CRLB, implying the ability to do better estimation. It has more recently been used as a design tool in, e.g., designing illumination patterns \cite{Balzarotti2017, Zhang2021} and optimizing the design of the tracker system \cite{Vickers2021}, as well as a means of comparing RT-FD-SPT performance \cite{VanHeerden2021}. 

\begin{figure}[htbp!]
    \centering
    \begin{subfigure}[t]{0.47\textwidth}
        \caption{Difference-over-sum estimator}
        \includegraphics[width=\textwidth]{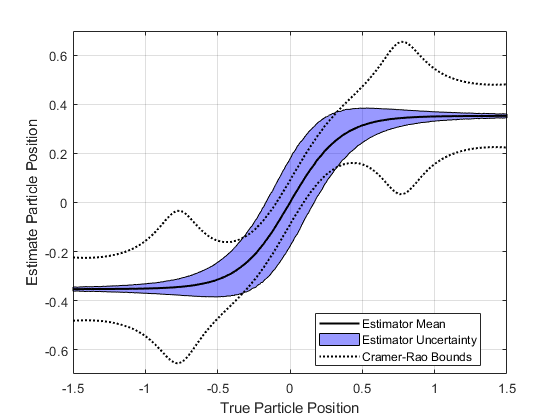}
        \label{fig:Estimator} 
    \end{subfigure}
    \begin{subfigure}[t]{0.47\textwidth}
        \caption{Localization Cramér-Rao Bounds}
        \includegraphics[width=\textwidth]{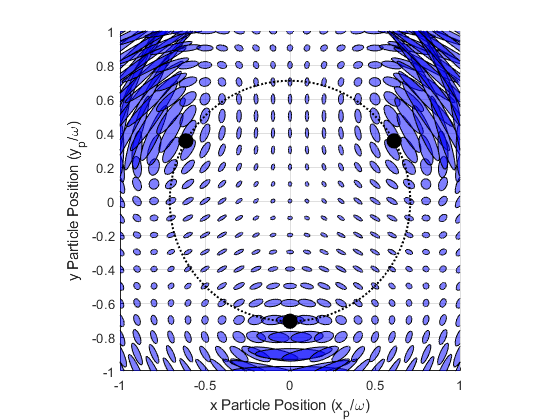}
        \label{fig:CrbMap} 
    \end{subfigure}
    \caption{Understanding the CRLB. (a) Results of 10,000 simulations of difference-over-sum estimation as a function of particle position. Black line: estimator mean, showing clear bias away from the center; purple region: $\pm3\sigma$ where $\sigma$ is the standard deviation in the estimates at each particle position; dotted black line: $\pm3\sqrt{CRLB}$. (b) The CRLB for the position of a particle in the plane using a 3-point measurement constellation (solid black dots), shown as an ellipse on a grid of particle locations. The major and minor axes of each ellipse are given by the largest and smallest eigenvalues, respectively, of the inverse of the FIM and demonstrate the dependence of the CRLB on the parameter values (see Supporting Information for more details). In both figures, distances are normalized to the beam diameter $\omega$ and are thus unitless.}
    \label{fig:CRB}
\end{figure}
When using the CRLB, it is important to verify the assumption that the estimator is unbiased. As an example, consider estimating the position of a particle along a line from measurements of the intensity using a simple difference-over-sum estimator. In \textbf{Fig.~\ref{fig:Estimator}} we show the the results of 10,000 Monte Carlo runs of such an estimator (see Supporting Information for more details). The plot shows the estimator mean (solid black line), the realized estimator uncertainty ($\pm3\sigma$, purple region), and the theoretical lower bound ($\pm3\sqrt{\textrm{CRLB}}$, dotted black line). Near the center, where the mean follows the true position (indicating that the estimator is unbiased), the estimator uncertainty is larger than the CRLB, as expected. At the edges, however, there is a clear bias in the estimator as it approaches a constant value. As a result, the uncertainty drops to zero and the CRLB is no longer a relevant guide.

It is also important to understand how the FIM changes with the true value of the parameters. To illustrate this, \textbf{Fig.~\ref{fig:CrbMap}} shows a visualization of the CRLB for estimating particle location as a function of particle position in the plane based on a 3-point measurement constellation (see Supporting Information for more details). The major and minor axes of each ellipse are defined by the maximum and minimum eigenvalues, respectively, of the inverse of the FIM given that the true particle location is at that point. Note that smaller ellipses indicate a lower CRLB and thus (possibly) increased precision in the estimate. Behind each measurement point the achievable precision is quite poor. The CRLB improves as the particle moves to points farther from, and symmetric to, the measurement locations. Such visualization can be immensely helpful when designing or evaluating a particular RT-FD-SPT system \cite{VanHeerden2021, Vickers2021}.

\section{Calibration and Characterization}\label{sec:calibration}

In order to maximize the performance of a given RT-FD-SPT system, and to understand its limitations, it is vital that the system is calibrated and characterized. Calibration is, of course, a natural component of the workflow in optical microscopy and is typically focused on understanding the actual PSF of the instrument and on determining the spatial resolution. The dynamic nature of RT-FD-SPT requires a more complex characterization that includes the additional steps of calibrating the actuation axes with respect to the optical coordinate system, finding the static localization precision, and determining the dynamic tracking performance. At each of these steps, adjustments to the hardware and tuning of algorithm parameters may be made; these changes often impact other elements of the systems and thus calibration and characterization is typically done in an iterative fashion as a system is tuned for a particular experimental aim.

\begin{figure}
    \centering
    \begin{subfigure}[t]{0.8\textwidth}
        \includegraphics[width=\textwidth]{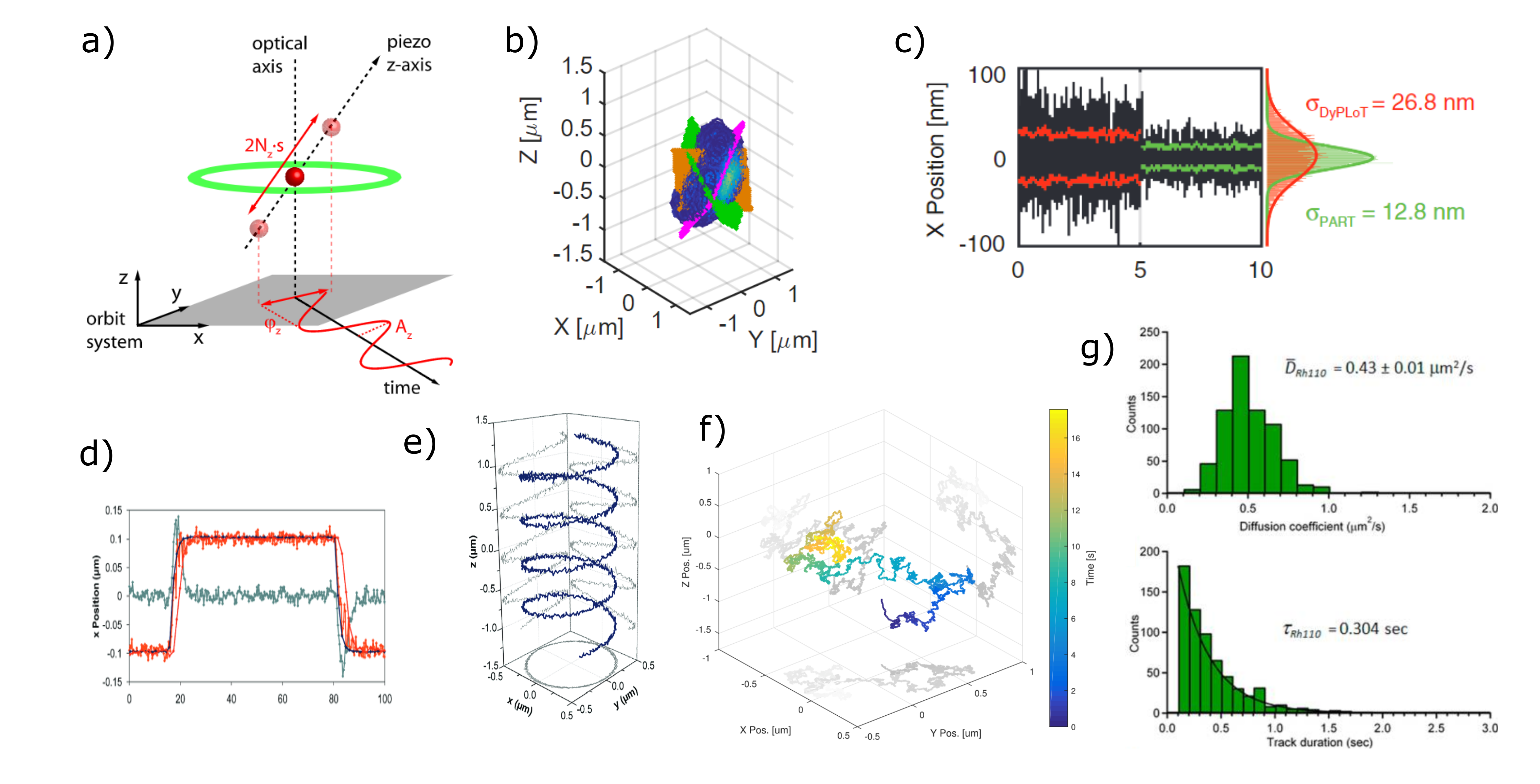}
        \phantomcaption
        \label{fig:CharacterizationAxes}
    \end{subfigure}
    \begin{subfigure}[t]{0\textwidth}
        \includegraphics[width=\textwidth]{Figures/Calibration2.png}
        \phantomcaption
        \label{fig:CharacterizationPSF}
    \end{subfigure}
    \begin{subfigure}[t]{0\textwidth}
        \includegraphics[width=\textwidth]{Figures/Calibration2.png}
        \phantomcaption
        \label{fig:CharacterizationStatic}
    \end{subfigure}
    \begin{subfigure}[t]{0\textwidth}
        \includegraphics[width=\textwidth]{Figures/Calibration2.png}
        \phantomcaption
        \label{fig:CharacterizationStep}
    \end{subfigure}
    \begin{subfigure}[t]{0\textwidth}
        \includegraphics[width=\textwidth]{Figures/Calibration2.png}
        \phantomcaption
        \label{fig:CharacterizationSpiral}
    \end{subfigure}
    \begin{subfigure}[t]{0\textwidth}
        \includegraphics[width=\textwidth]{Figures/Calibration2.png}
        \phantomcaption
        \label{fig:CharacterizationStokesEinstein}
    \end{subfigure}
    \begin{subfigure}[t]{0\textwidth}
        \includegraphics[width=\textwidth]{Figures/Calibration2.png}
        \phantomcaption
        \label{fig:CharacterizationDDistribution}
    \end{subfigure}
    \caption{Examples of calibration and characterization. \textbf{(a)} Characterization of actuator to optical axes alignment (adapted with permission from~\cite{Ernst2012} © The Optical Society). \textbf{(b)} PSF characterization showing rotation with respect to actuator axes (adapted with permission from~\cite{Ashley2016a} © The Optical Society). \textbf{(c)} Fluctuations of tracking system to a static particle (reprinted from~\cite{Hou2019}. © 2019 Wiley-VCH). \textbf{(d)} Step response of system (adapted with permission from~\cite{Juette2010} © 2010 American Chemical Society). \textbf{(e)} Tracking helical trajectory (adapted with permission from~\cite{Juette2010} © 2010 American Chemical Society). \textbf{(f)} Tracking free diffusing particle (adapted with permission from~\cite{Ashley2016a} © The Optical Society). \textbf{(g)} Distribution of measured diffusion coefficients (top) and tracking duration of freely diffusing particles (bottom)  (adapted with permission from~\cite{Han2012} © 2012 American Chemical Society).}
    \label{fig:Calibration}
\end{figure}

%PSF
Characterizing the PSF of the specific system being used is as important in RT-FD-SPT as in standard optical microscopy. Accurate knowledge of its shape can help in selecting  algorithm parameters, such as the orbital radius for single-detector scanning methods \cite{Berglund2006, Gallatin2012, Vickers2021}, and is a key element when determining the minimum uncertainty achievable in particle position estimation \cite{ram2006stochastic, Gallatin2012, chao2016fisher, Vickers2021}. There are well-known techniques for PSF characterization (see, e.g., \cite{Juskatis2006}) with the essential idea being to fix a small fluorescent particle like a fluorescent nanosphere or a quantum dot to a coverslip, acquire fluorescence intensity measurements from a 3D raster scan of the excitation light beam across the particle, and fit this data to an experimental model. It is important that the dimensions of the fluorescent particle are significantly smaller than those of the focused light beam so that it may be considered a point source of fluorescence that has a negligible effect on the measured PSF of the optical system. To avoid the effect of the coverslip on the axial component of the PSF, the fluorescent particle is typically embedded in a polymer or gel when characterizing the axial component of the PSF \cite{petrov2020addressing}. Alternatively, assuming the PSF to be symmetric about the XY plane, a substrate-bound particle can be used to determine the axial component only in the direction of the laser beam (away from the coverslip). To limit spherical aberration or broadening of the laser focus due to refractive index mismatch between the immersion medium and the sample, it is advisable to use a water-immersion or water-dipping objective.

In general, the optical axis of a RT-FD-SPT system will not be exactly aligned with the vertical axis of the actuators. This misalignment creates systematic errors in estimation if it is not known and can lead to vertical motion being misinterpreted as dynamics in the lateral directions \cite{Ernst2012}. An example of this is shown in  \textbf{Figure~\ref{fig:CharacterizationAxes}} where this effect was demonstrated on an orbital tracking RT-FD-SPT microscope \cite{Ernst2012}. This misorientation can be easily determined from the PSF characterization data and represented as a simple 3D rotation \cite{Ashley2016a} (\textbf{Figure~\ref{fig:CharacterizationPSF}}). Alternatively, if the beam and a sample stage can be scanned independently, one could use a fluorescent particle and move it along each actuator axis while doing a 3D raster scan, an orbital scan, or signal maximization. As this calibration need is unique to tracking microscopes, it represents an area requiring further development.

Once the PSF has been characterized, the static localization performance of the system can be investigated. This performance is affected by several factors, including the PSF of the instrument, the ability of the system to lock onto a fixed particle, and the steady-state behavior of the algorithm when ``tracking'' a fixed particle (\textbf{Figure~\ref{fig:CharacterizationStatic}}). These elements can be characterized using the same physical setup as that for characterizing the PSF, namely a particle fixed to a coverslip (or embedded in a polymer or gel). Using this method, Hou and Welsher showed that the controller gains have a large effect on the size of these fluctuations \cite{Hou2017}, emphasizing the need for good tuning of controller parameters. Note that there is a balance to be achieved here: selecting gains that are too aggressive can lead to large fluctuations that can increase the variance of the position estimates or even result in loss of tracking, while selecting them too conservatively can lead to lack of sensitivity and an inability to following moving particles. 

The final piece in calibration and characterization is studying the tracker performance in dynamic tests. These tests can be collected into four categories that make different trade-offs when weighing the need to know ground truth against the biological relevance of the test conditions: deterministic motion, simple diffusion, engineered samples, and synthetic motion. In deterministic motion, a particle is once again fixed to a coverslip. The coverslip is then moved in a known and controlled way using, e.g., a nanopositioning stage. Piezoelectric-based nanopositioning stages are tested and calibrated by the equipment manufacturers using high precision laser interferometers and typically provide nanometer or better positioning accuracy and precision, making them excellent choices for generating the (known) reference motion. Typical patterns include steps \cite{Gratton2003,Cang2006, Juette2008,Juette2013,Annibale2015,Zhang2021} ((\textbf{Figure~\ref{fig:CharacterizationStep}}), circles \cite{Juette2010}, sinusoids \cite{Gratton2003, Gratton2005, Katayama2009, Hou2017, Balzarotti2017}, and helices \cite{Juette2010, Annibale2015, Perillo2015} (\textbf{Figure~\ref{fig:CharacterizationSpiral}}). Because the ground truth of the motion is known, this approach can be used to characterize response time and tracking error. However, these methods are not representative of the motion of a particle in an aqueous or cellular environment and do not allow for the study of performance with respect to motion parameters such as diffusion coefficients.

Simple diffusion is a well-accepted model for describing the dynamics of biological macromolecules in their native environments. In this second category, particles of known size are placed in an aqueous environment of known viscosity (\textbf{Figure~\ref{fig:CharacterizationStokesEinstein}}). The diffusion coefficient, governed by the well-known Stokes--Einstein relation, can be selected by controlling the particle size, the viscosity of the solution, and the temperature. This method has been performed using fluorescent microspheres \cite{Cang2007,Ernst2012, Dupont:2013bu, Juette2013, Lien2014, Lanzano2014, Gremann2014, Hou2019}, ``giant" quantum dots \cite{DeVore2015a, Hou2017}, and regular quantum dots \cite{Lessard2007, McHale2007, Wells2008, Shen2011, Shen2011a, Ashley2012, Welsher2014, Ashley2016a} diffusing in solutions of agarose \cite{Gratton2003, Ashley2016a}, sugar-water \cite{Ragan2006, Dupont:2013bu}, glycerol-water \cite{Gratton2005, Cang2006, Lessard2007, Cang2007, Wells2008, Shen2009a, Hellriegel2009, Juette2010, Shen2011, Shen2011a, Ashley2012, Han2012, Ernst2012, Ernst2013, Lien2014, Gremann2014, Perillo2015, Ashley2016a, Hou2020}, and water \cite{McHale2007, Hou2017, Hou2019, Hou2020}. After a tracking experiment, the diffusion coefficient and localization error can be estimated from the localizations using, e.g., a mean squared displacement (MSD) analysis \cite{michalet2012optimal} or a Maximum Likelihood Estimator (MLE) \cite{Berglund:2010ffa} and statistics on performance gathered by repeating multiple experiments (\textbf{Figure~\ref{fig:CharacterizationDDistribution}}). While this motion is biologically relevant, there is no ground truth on the particle trajectory and even the ``known'' diffusion coefficient is inexact due to factors that include dispersion in the sizes of manufactured particles and local variations in the viscosity and temperature of the sample. 

The third category is similar to simple diffusion but rather than using a simple aqueous solution, the particles are placed in an engineered environment that increases the complexity of the particle dynamics \cite{saxton2012wanted}. Researchers have used oil--water interfaces \cite{Du2012}, and  proposed engineered lipid membranes, mesoporous materials, opals, and polymer solutions \cite{saxton2012wanted}. While this approach is more difficult to set up, the particle motion may be more representative of a situation of interest. As with simple diffusion, however, ground truth is essentially impossible to come by.

The final category is a method called synthetic motion \cite{saxton2012wanted, Vickers2020feedforward, vickers2022synthetic}. As with the method of deterministic motion, this approach uses a fluorescent particle fixed to a coverslip and moved by a nanopositioning stage. Instead of a simple, deterministic trajectory, however, the stage is used to realize almost any biologically relevant process, including simple diffusion, tethered motion, confined diffusion, hop diffusion, and more. Because these motions are realized artificially, ground truth is known on trajectory and motion model parameters. However, the local environment remains simple, without confounding factors such as background fluorescence from nearby structures, and creating accurate realizations of the desired motion, particularly for fast dynamics, can be challenging and is ultimately limited by the capabilities of the nanopositioning stage being used~\cite{vickers2022synthetic}.

As RT-FD-SPT microscopes add multi-color detection, contextual imaging, and spectroscopy, it is important to calibrate and characterize each additional function and to ensure the co-alignment of each element in the system. A major challenge is that of unintentional motion resulting from thermal expansion, stage drift, and other sources. While the tracking controller is robust against these unintentional motions and will remain locked onto the particle of interest, these unintentional motion sources will affect any position or dynamic parameter estimation. To mitigate this and to achieve nanometer precision over long durations, the stage holding the sample can be stabilized with active feedback. This can be achieved in multiple ways including imaging of gold nanorods \cite{Schmidt2021}, using fiducials \cite{coelho20213d}, and image correlation \cite{mcgorty2013active}, and commercially available systems that use back-reflection of infrared light off of the coverslip, to provide a signal for the control to hold steady, thereby providing a stable global frame of reference. Aligning spectrally separated detection channels faces the same challenges as in other microscopes, and commercially available fluorescent microspheres with multiple different fluorescent dyes can be used for chromatic registration. Similarly, NIST traceable fluorescent samples enable calibration for fluorescence intensity and wavelength \cite{gaigalas2001development}. Finally, it is generally useful to obtain a contextual image simultaneously with the tracking. To register these two data streams, one must learn a  transformation between the camera and tracking channel coordinates. This can be achieved by imaging with both channels a sample that has multiple fluorescent beads across the entire field of view, and then fitting a transformation between both images.

\section{Tracking Performance}\label{sec:Performance}
The performance of RT-FD-SPT is a challenging topic to cover. This is due to the stochastic nature of RT-FD-SPT as well as inherent trade-offs between different aspects of performance.  There is no single metric that completely describes the behavior of RT-FD-SPT systems, and it is important to consider these techniques with respect to many different factors, each of which may have different relevance in different applications. Additionally, reporting of system performance has not been consistent, and existing methods have not necessarily been pushed to their limits, making direct comparisons between different systems difficult. In this section, we first cover the currently reported performance metrics.  Following this we describe a set of practical performance metrics with the goal of guiding a researcher interested in applying RT-FD-SPT in selecting an approach. One of the challenges is knowing the ultimate performance of a system. A promising approach is to analyze the system performance from the fundamental statistics of the tracking process and the resulting trade-offs. Finally, we discuss performance fit to experimental aims.

\subsection{Current Performance}
%survey how performance is currently reported and the results of it
Several reported metrics have been used to quantify the performance of RT-FD-SPT methods, including localization precision, fastest tracked diffusion coefficient, longest tracking duration, lowest photon count rate, and experimental tracking precision. These can be lumped into two broad categories: tracking precision and tracking success. 

For precision, various theoretical metrics have been used. The Fisher information and CRLB provide an indicator of the precision with which a static particle can be localized, typically as a function of the number of detected photons~\cite{Gallatin2012,Balzarotti2017,VanHeerden2021,Vickers2021,Zhang2021}. This is useful since ideally the feedback is fast enough that the particle can be considered to be approximately static during one cycle. Another performance metric is the mean-squared tracking error, which can be predicted theoretically using suitable approximations~\cite{Berglund2006} or using a numerical simulation~\cite{Berglund2004,Andersson2005,Berglund2006,Andersson2009,Wang2010,Fields2012}. The precision of an RT-FD-SPT system depends on many factors. The CRLB is of course influenced strongly by the number of detected photons and the signal-to-background ratio, but also by the way in which the position is measured. Optimal measurements can have a significant impact on the precision~\cite{Gallatin2012,Balzarotti2017,VanHeerden2021,Zhang2021}. Also, in general, a decrease in the detection volume size corresponds to a lower CRLB (and thus increased precision). The simulated mean tracking error is dependent on the CRLB but also depends on the estimation and control algorithms and the system model.  

%While these metrics focus on the precision of the tracking, i.e., how accurate the tracking is
Metrics that focus on the ``success" of the tracking include the expected tracking duration~\cite{Enderlein2000a}, the percentage of particles lost within the simulation run time~\cite{Fields2012}, and the largest diffusion coefficient that can consistently be obtained for some length of time~\cite{Berglund2004,Andersson2005,Andersson2009,Fields2012,VanHeerden2021}. The tracking success strongly depends on the feedback bandwidth of the overall tracking system, the size of the detection volume, and the rate of photon detection. As one would expect, broader bandwidths and larger detection volumes allow for faster particles to be tracked and for longer duration. In contrast, higher precision requires a smaller detection volume~\cite{VanHeerden2021}. There is, therefore, generally a tradeoff between the precision and success of tracking. For this reason, one cannot simply combine these metrics into a single performance metric, as different applications might favor either speed or precision.

\begin{figure}[t]
    \centering
    \includegraphics[width=0.8\textwidth]{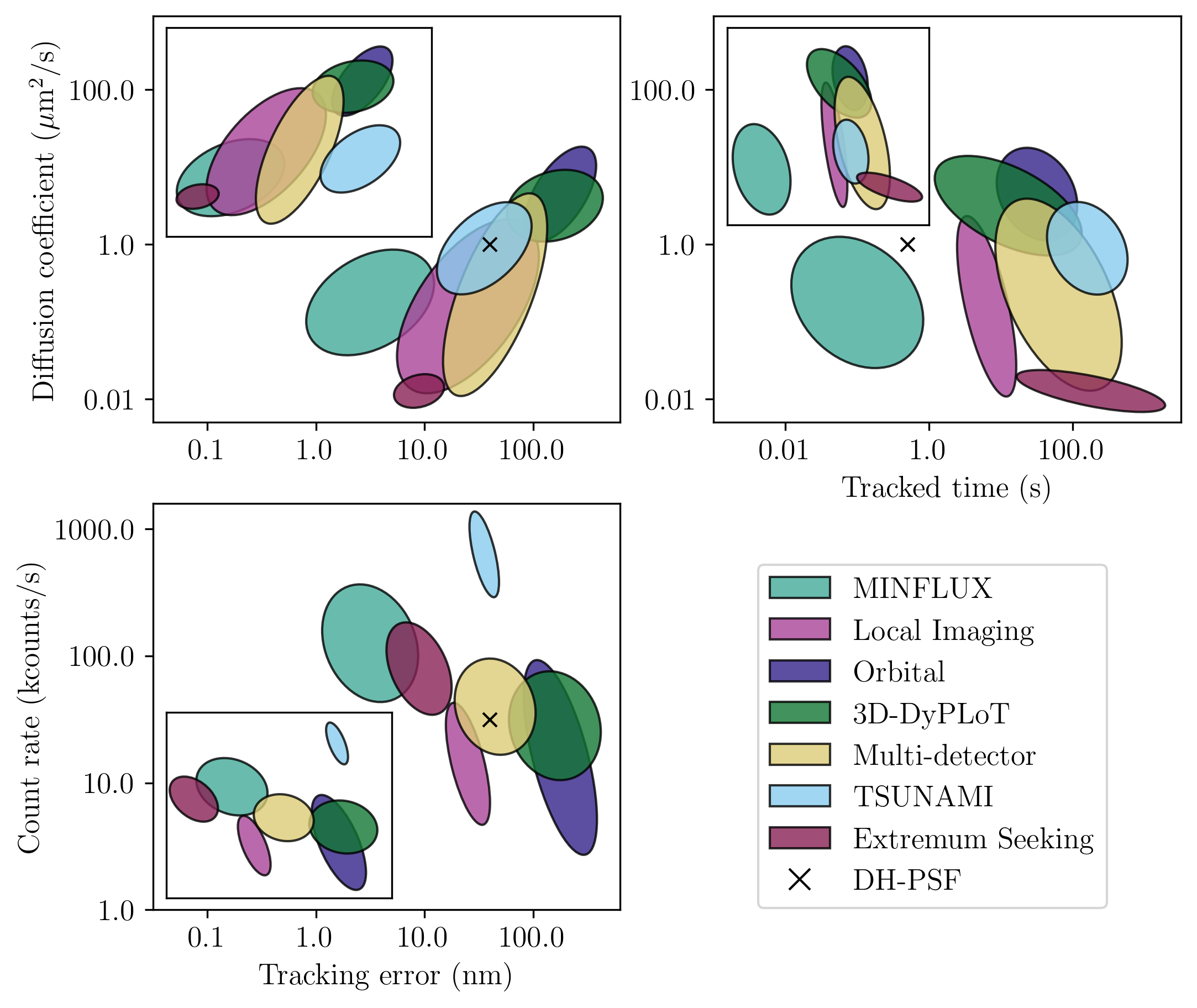}
    \caption{Performance comparison of different RT-FD-SPT methods. Insets show cost-adjusted values, which were calculated by dividing the values of the diffusion coefficient, tracking error and tracked time by the cost in Table~\ref{tab:cost}. As the photon count rate depends more critically on the sample used than anything else, it was not normalized to the cost. Values are based on the ranges of reported values from the following references: MINFLUX:~\cite{Balzarotti2017} Local Imaging:~\cite{Juette2010}, Orbital 3D Highspeed:~\cite{McHale2007,Berglund2007a,Berglund2007b,McHale2009,McHale2010}, 3D-DyPLoT:~\cite{Hou2017,Hou2020}, Multi-detector:~\cite{Welsher2014,Wells2010}, TSUNAMI:~\cite{Perillo2015}, Extremum Seeking:~\cite{Ashley2016a}, double-helix point spread function (DH-PSF):~\cite{Carr2017}.}
    \label{fig:performance}
\end{figure}

To compare different methods' experimental performance, it is necessary to consider the interplay between different metrics. In \textbf{Figure~\ref{fig:performance}} we show the performance of different approaches with respect to four key metrics: diffusion coefficient, tracking error, tracked time, and photon count rate. Instead of simply considering the best reported values, we based the comparison on ranges of reported values to illustrate the regions of parameter space where the methods are known to work well (see Supporting Information for details). It is possible that each method can be applied outside of its known region, but such results have not yet been demonstrated, to our knowledge. There are, of course, also other metrics that could be considered, for example tracking depth, maximum tracking range, or metrics related to spectroscopic measurements such as fluorescence lifetimes (see \textbf{Section~\ref{sec:ApplicationFit}}). The ones shown here are what we consider to be the most generally applicable. 

As a reference point, we also show one example of the reported performance of one conventional image-based 3D-SPT technique, namely the use of the double-helix point spread function (DH-PSF)~\cite{Carr2017}. We stress that this is simply a reference and not meant to represent the best performance of this technique, and this review does not intend to compare the performance of RT-FD-SPT and conventional image-based SPT approaches. The metrics shown here are not the ones most commonly used to evaluate conventional SPT performance and therefore are often not all reported, making a direct comparison between image-based and RT-FD-SPT performance non-trivial. This is, however, a promising avenue for further research. In any case, as discussed in \textbf{Section~\ref{sec:Survey}}, image-based and RT-FD-SPT are better thought of as complementary rather than competitive techniques.  

As expected, in all metrics, all techniques show improved performance under easier conditions (i.e., slower diffusion values, higher count rates, etc.). Because cost is also a significant factor to take into account when selecting a method, the inset plots in Figure~\ref{fig:performance} show the performances normalized by the hardware cost shown in \textbf{Table~\ref{tab:cost}}. These calculated costs are based on the most expensive components in the experimental realizations of the methods, such as the laser source, photodetectors, control hardware and software, and beam scanning components  (details on these cost estimates can be found in the Supporting Information). We note that it may be possible to achieve comparable performance at a lower cost through careful choice and optimization of equipment. 

There are a few clear trends that emerge in Figure~\ref{fig:performance}. For example, the comparison between the tracked diffusion coefficient and tracking error clearly indicates the trade-off between tracking speed and localization performance. While different techniques approach this trade-off in different ways, these results indicate you must give up on one to gain in the other. As mentioned above, this is the main reason why these metrics cannot simply be combined into a single, general measure of performance. Similarly, there is a clear decrease in tracking error with increasing photon  count rate in all methods. Interestingly, while tracking time generally increases with decreasing diffusion coefficient, there is less of a trend in this comparison, possibly reflecting that factors other than loss of tracking led to the end of an experimental run. Comparison of the methods shows that those that implement fast beam scanning (MINFLUX, orbital scanning, and 3D-DyPLoT) have the better reported performance, but also have high cost and complexity. MINFLUX has a clear edge in precision, but lags when it comes to tracking duration. TSUNAMI has good performance, but appears to depend on very high photon count rates. TSUNAMI and MINFLUX are, however, significantly more expensive than the other methods (in the case of TSUNAMI this is due to the use of a femtosecond laser) and thus their position worsens in the cost-adjusted plots. Conversely, the extremum seeking method's minimalistic hardware requirements make it inexpensive, and thus it has excellent performance-per-dollar. 

% \begin{table}
%  \caption{Table 1 caption}
%   \begin{tabular}[htbp]{@{}lll@{}}
%     \hline
%     Description 1 & Description 2 & Description 3 \\
%     \hline
%     Row 1, Col 1  & Row 1, Col 2  & Row 1, Col 3  \\
%     Row 2, Col 1  & Row 2, Col 2  & Row 2, Col 3  \\
%     \hline
%   \end{tabular}
% \end{table}

\begin{table}[htbp!]
    \caption{Approximate cost of 3D RT-FD-SPT experimental setups in US Dollars}
    \centering
    \begin{tabular}[H!]{@{}lll@{}}
        \hline
        Method  &  Cost (USD)\\
        \hline
        TSUNAMI & 195k\\
        MINFLUX 3D & 170k\\
        Optical Cantilever & 97k\\
        3D-DyPlot & 95k\\
        Orbital 3D Highspeed & 83k\\     
        Tetrahedral & 83k\\       
        Quadrant & 82k \\
        Local Imaging & 82k\\
        Orbital 3D with tunable lens & 55k\\
        Extremum Seeking & 55k\\
    \end{tabular}
    \label{tab:cost}
\end{table}

\subsection{Fundamental Performance and Trade-offs}
The characterization of RT-FD-SPT performance has historically focused on practical and measurable performance.  This is very useful but can make it difficult to compare systems, as each system needs to be tested in the same way and to the limits of its performance, which can be time-consuming and offers no guarantee that the best performing configuration of that particular system will be found.  Additionally, experimentally measuring the performance does not provide a systematic way to design and improve these systems. A different approach is to look for fundamental concepts that underlay and unify RT-FD-SPT performance, and explain the experimentally observed trends in localization precision, tracking error, and tracking duration. The two fundamental ideas underlying these performance metrics are trackability and noise propagation. 

The aim of quantifying trackability is to understand how different aspects of the system enable or inhibit locking onto and tracking the particle for an extended duration. This can be seen as the theoretical counterpart to tracking success where there is a correspondence with both the experimentally derived maximum tracked diffusion coefficient and maximum tracked duration discussed in the previous section. The analysis of trackability has two equivalent points of view. The first notion of trackability focused on understanding the probability of not losing a particle for a finite duration and was analyzed for the cases of a particle escaping the detection volume \cite{Andersson2005}, for a particle going into a dark state \cite{Wells2008}, and for photobleaching \cite{Han2012}. Using this framework, a maximum trackable diffusion coefficient was calculated \cite{Andersson2005}. The argument of this tracking probability gives a dimensionless number, $\frac{r}{\sqrt{DT}}$, where $r$ is the characteristic dimension of the localizable volume, $D$ is the diffusion coefficient, and $T$ is the integration time. Larger values of the dimensionless number indicate better trackability. It was then shown that tracking duration follows a geometric distribution related to the tracking probability \cite{Wells2008}. The tracking probability directly relates to the mean duration of tracking. The second approach focuses on understanding the probability distribution of the time for a particle to first exit the detection volume given a fixed detection volume. The first moment of this distribution gives us %A second approach uses 
the mean escape time \cite{Gremann2014, Ashley2018}, which is the average time to lose a particle,  %time at which the particle first exits the detection volume. 
%The escape time 
and is proportional to $\frac{r^2}{D}$. An effective controller increases the tracking duration, which increases the value of the proportionality constant \cite{Ashley2018}. A missing aspect of these analyses is the effect of photon detection. By considering both particle motion and photon detection, Berglund developed the dimensionless number $\frac{r^2 \Gamma}{D}$ as a metric for trackability, where $\Gamma$ is the mean detected photon rate \cite{Berglund2007}. Although research in this area is far from over, these results provide a basis for improving the duration of tracking in a systematic way. 

The second fundamental aspect of performance is how noise propagates through the system.  Both the motion of the molecule and the detection of photons are fundamentally stochastic processes and they are typically the dominant sources of noise. This has been characterized in two key ways.  The first was the adoption of the Fisher information and CRLB \cite{Gallatin2012, chao2016fisher, Balzarotti2017, VanHeerden2021, Vickers2021}. These provide a basis of understanding the best possible precision with which a particle can be localized as well as a lower bound on the noise from photon detection and online estimation of position.  Using these metrics as a tool for analysis, it was shown that smaller beam widths and a larger number of detected photons give better localization precision \cite{Gallatin2012, Vickers2021}, essentially reducing the impact of stochastic variation in the photon signal. Additionally, different laser modes can give more precise localizations \cite{Balzarotti2017}. The second analysis approach uses the lens of stochastic dynamic systems to understand how diffusion and shot noise limit system performance \cite{Berglund2006,Berglund2007, Berglund2007a}. It was shown that tracking error follows a $\sqrt{D}$ trend. Additionally, there is a minimum tracking error that is proportional to localization uncertainty and the bandwidth of tracking and online estimation. Further, the photon detection rate sets a lower limit on the tracking error variance.

%tradeoffs
One of the most frustrating and interesting aspects of RT-FD-SPT is the emergence of parameters that are subject to a trade-off between localization uncertainty and tracking duration. The first parameter to consider is that of photobleaching. Photobleaching puts a limit on the expected number of photons one can use for the competing objectives of tracking and localization. Increasing laser power improves localization at each time point, but it also increases the probability of bleaching, which reduces tracking duration. Another parameter to consider is that of the beam diameter and scan pattern. It was shown that smaller beam diameters (and thus a smaller detection volume) improve localization uncertainty \cite{Gallatin2012, Vickers2021} while increasing the detection beam diameter and volume increases the maximum trackable diffusion coefficient \cite{Andersson2005, Berglund2006}, enabling locking onto fast diffusing particles for longer duration \cite{McHale2007, Hou2019}. One interesting route out of this trade-off stalemate comes from adaptation of the beam diameter or scan pattern in response to the particle's diffusion coefficient, providing better localization for slow diffusing particles without sacrificing tracking duration of fast diffusing particles~\cite{Hou2019}. The final trade-off parameter is the system bandwidth \cite{Berglund2006, Berglund2007}. Long integration periods give time to collect more photons for localizing the particle, but also increase the likelihood of losing the particle, implying that there is an optimal bandwidth or integration time for a given weighting between these two competing phenomena.

\subsection{Performance--Application Fit}\label{sec:ApplicationFit}
Much of the testing and analysis of RT-FD-SPT systems was done in the context of determining a particle's trajectory and inferring its dynamic model. In this experimental aim, both tracking duration and localization precision are important in determining the uncertainty in the estimates of the model parameters. However, an optimal position measurement does not, in general, correspond to optimal inference of model parameters. Similarly, in spectroscopy applications (Section~\ref{sec:Spectroscopy}), an optimal position measurement does not generally correspond to an optimal spectroscopic measurement. Determining the best approach for a specific application in general involves assessing four factors: the speed (e.g., the diffusion coefficient) of the particle of interest, the achievable photon detection rate under experimental constraints, the time scale of the process under investigation, and, if a separate beam is used for concurrent spectroscopy, the detection volume of this beam. Intuitively, the tracking error of a technique should be smaller than the  detection volume of the spectroscopic beam to ensure uninterrupted and quality data. 
%Determining the most appropriate technique for a particular application requires assessment of how the currently reported performance shown in Figure \ref{fig:performance} indicates fit to the specific question and system of interest. In general, there are four factors Assessing this fit in general involves understanding (at least) four factors of their experiment.  These are: the speed of the particle (eg. diffusion coefficient), the photon detection rate, the time scale of the process under investigation, and detection volume of the spectroscopic beam.  Intuitively, tracking error is important as this should be less than the spectroscopic detection volume.  
Additionally, experiments with slow processes or low detected spectroscopic photon rates will require longer duration tracking to attain a particular measurement uncertainty, while fast moving particles will need a high temporal resolution to maintain the particle in the detection volume of the tracking beam and, if used, of the spectroscopic beam. Finally, when long tracking durations are needed, the system will generally need a larger tracking range to follow the particle through a larger volume. With these considerations, selection of a technique can be guided at least in part by reported performance (Figure~\ref{fig:performance}).

While these considerations help to guide a selection of technique, they are qualitative, and a formal and quantified approach to evaluating performance and fit to a specific application remains an unexplored aspect of RT-FD-SPT spectroscopy. This represents an open challenge for future development. A promising approach is to develop expressions for the Fisher information and CRLB for a particular experimental aim.  This has been done in the case of FRET and charge-transfer techniques \cite{Watkins2004, Watkins2005}. When this is done in the context of RT-FD-SPT, a connection between experimental aim, relevant system parameters, and performance metrics is established, enabling an assessment of application-specific performance. A design that optimizes this relationship may end up being different from the optimal design for position measurement. %To our knowledge, this fact has not been explored yet, but should be kept in mind by anyone interested in combining RT-FD-SPT and spectroscopy. 

%\begin{figure}
%    \centering
%    \includegraphics[width=5.5in]{Figures/ExperimentalFit.png}
%    \caption{Performance to application fit for applications listed in Figure 4. Longer bar length indicates greater importance of the metric.}
%    \label{fig:ApplicationFit}
%\end{figure}

\section{Future Perspectives}
\label{sec:outlook}

As demonstrated in this review, RT-FD-SPT research has been productive, leading to several different methods, and clearly demonstrating the potential of these techniques for biological research. In this section, we discuss our envisioned future of the field with regard to research into both the technique itself and applications. We remark on the trends of RT-FD-SPT, development challenges and opportunities, and the adoption in the wider scientific community.

\subsection{Current and Future Trends}
\label{sec:curr_fut}
Research into RT-FD-SPT methods continues and recent work shows several promising trends. One example is an improvement in performance over time, especially with regard to tracking error (see Figure~\ref{fig:performance}). A few very recent studies indicate interest in, and demonstrate the importance of, developing a theoretical understanding of RT-FD-SPT, particularly with regard to statistical analysis of precision ~\cite{VanHeerden2021,Vickers2021,Tan2021}. Similarly, developing a fundamental understanding of the elements that determine how long particles can be tracked has led to a general increase in the duration of tracking. As our knowledge of the fundamental trade-offs between tracking precision and tracking duration continues to improve, researchers are designing their systems with these trade-offs in mind. This has enabled longer tracking duration with only minor increases in position estimation uncertainty. These recent developments that leverage the statistical basis of performance are expected to grow and to extend to other important aspects of RT-FD-SPT such as control design and tailoring a system for a specific experimental aim.

There are also promising developments in the controllers applied in RT-FD-SPT. Controller design has begun to move away from simple, PID-based tracking controllers toward more advanced techniques such as optimal, nonlinear, and adaptive control. Two key drivers for these trends are that the problem of tracking a diffusing particle is fundamentally a stochastic one, and that the experimental context changes over time. The next generation of RT-FD-SPT methods will need to strike a balance between enhancing performance through advanced control techniques, and ensuring ease of implementation to promote adoption. This tension in that balance provides unique challenges and opportunities for designers.

Another set of trends in RT-FD-SPT is advances in the design of the excitation and detection sub-systems. These trends can be broken down into increasing the tracking performance of the system, understanding the experimental context, and overcoming the limitations of fluorescence-based tracking. A promising direction looks at real-time optimization of the laser illumination intensity to improve the performance of the system.  This was first done by McHale \textit{et al.}~\cite{McHale2007} for the tracking duration, and by Balzarotti \textit{et al.}~\cite{Balzarotti2017} in MINFLUX. This trend of adjusting or engineering the illumination to meet the experimental needs is likely to continue and to produce additional performance gains. The engineering and optimization of the illumination pattern and detection is a largely unaddressed opportunity. 

As experimental aims move beyond just tracking the particle, complementary information must also be acquired. This is first seen as the trend toward the inclusion of various spectroscopic methods to infer more than just the particle's position and dynamics, measuring a more complete state of the particle. Some of the emerging experimental aims require estimating the particle's relative position to some other particle or feature. This could be the folding of proteins or binding onto a specific receptor. To meet these needs, there has been a trend of adding additional detection channels. One form of this is the addition of multi-color detection allowing for localization of two separate fluorescent labels \cite{Liu2020a, Keller2018}. Some of the emerging experimental aims include understanding the particle's behavior in the context of the cellular environment such as proximity to the cell membrane. To meet this need, there has been research focusing on including simultaneous imaging of the cellular environment %experimental context 
\cite{Welsher2014, Johnson2019}. The trend of adding parallel co-aligned measurements and imaging modalities will only continue as the experimental aims move beyond observing the motions of a single particle.

One trend we may see is the development of tracking methods that do not use fluorescence such as interferometric scattering (iSCAT)~\cite{Taylor2019,Taylor2019a}. iSCAT considers the interference between the light scattered by the particle of interest and the portion of incident light that is reflected by a nearby interface. It therefore represents a different source of photons than fluorescence. Although iSCAT has already proven to be a powerful imaging technique, it has only been used once with real-time feedback, in the form of the ABEL trap~\cite{Squires2019}. iSCAT overcomes many of the trade-offs of fluorescence-based tracking as discussed in Section~\ref{sec:Performance}, as saturation and photobleaching can be minimized by choosing a wavelength of light that is negligibly absorbed by the tracked particle. This makes it a very attractive technique for RT-FD-SPT. A recent study~\cite{VanHeerden2021} showed that iSCAT should perform better for some samples (particularly relatively large ones), even at the same illumination intensity. Seeing as a much higher illumination intensity ($\sim1000$ times higher) is typically used for iSCAT with no phototoxic effects, the method should, in fact, offer excellent performance for a wide range of samples. A major drawback of iSCAT, however, is that it is very challenging to use \textit{in vivo}. This is firstly due to the crowded molecular environment: as all molecules scatter light, iSCAT fails to provide the same selectivity offered by fluorescent labeling, giving rise to significantly enhanced background that severely compromises the signal-to-background ratio. Secondly, iSCAT typically requires a particle to be close to a reflective interface, as the method requires coherence between the scattered and reflected light fields.

\subsection{Open Challenges and Opportunities}
%open challenges and opportunities.
%controls
As the field of RT-FD-SPT pushes the boundaries of performance and application, open challenges and development opportunities emerge, both in individual components of the tracking system and in the combination of these components into a total system.  We start this discussion with the opportunities in controller design. To date, high-bandwidth tracking implements very expensive beam scanning and stage scanning actuators. There is a need to develop affordable high-bandwidth actuation systems through refining their design or through improved control. As bandwidth increases, the limiting factor for tracking will transition away from actuator limitations and toward photon detection, opening up opportunities in considering more complex tracking controllers.  We have seen that the application of adaptive, optimal, and nonlinear controllers improves overall tracking performance over traditional PID feedback approaches. Further development of adaptive control approaches to enable the system to react in real-time to changes in the motion of the tracked particle, changes in the experimental context, and changes in the detection of photons promises exciting advances in technique and application. While there have been some initial efforts in this area, there remains significant opportunities to use adaptation in the face of uncertainties and error arising from measurement noise, modeling error, motion blur, and other sources. The use of adaptation shifts the paradigm of RT-FD-SPT to greater levels of autonomy, enabling single molecule experiments to transition from single measurements to large data sets. Similarly, if the initialization of tracking could be automated and optimized, overall throughput and utility would be significantly enhanced.

%Estimation
Similar to controller design, there are opportunities and open challenges for estimation in RT-FD-SPT. To date, one of the main goals for online estimation was that it must be easy to implement and fast to compute. This has produced estimators that show increased error as the particle moves away from the center of the detection pattern. There is an opportunity to implement online position estimators that have lower error over a larger area of the detection pattern and to couple these estimators more directly to new, adaptive controllers. An additional limitation of current approaches is that the noise from photon detection is typically not filtered or taken into consideration in the control loop.  A few groups have used Kalman filters to help mitigate the impact of this noise; however, the adoption of filtering should be routine.  Another possible avenue for research concerns the photon statistics itself. Thus far, when photon noise has been taken into account in estimators for RT-FD-SPT, it has generally been assumed to be Poisson-distributed. However, single emitters exhibit sub-Poissonian noise~\cite{Mandel1995}. To achieve the ultimate performance in tracking single molecules at low photon count rates will likely require taking this into account. It is particularly important as RT-FD-SPT systems move toward single photon estimation and control where accurate noise models and filtering become exceedingly important. Finally, it would be of significant interest to compare the performance of RT-FD methods with that of more traditional, camera-based techniques in applications where the capabilities of the different methods overlap. Understanding and comparing how these different experimental techniques handle noise and estimation error, and how these errors affect the final parameter estimation, is important in evaluating methods when determining what is most appropriate to address an experimental aim.

%illumination and detection
The development of RT-FD-SPT microscopes has largely focused on using Gaussian lasers in combination with a scan pattern, or a set of detectors.  While a Gaussian laser is practical and ubiquitous, it is unclear if it is optimal for particle tracking. MINFLUX has demonstrated that other laser modes and specially engineered beam profiles can provide better localization precision \cite{Balzarotti2017}. The optimization of the laser illumination and detection for tracking performance or to fit the experimental aim is still an open challenge and questions related to understanding the benefits and weaknesses of alternate illumination approaches are largely unexplored. As research emerges to address this, it will be important to develop methods and performance metrics that allow researchers to understand how to select the illumination and detection patterns that best fit their application and experimental aim.  

%Performance
Finally, the quantification of total system performance is an area that is still evolving. Understanding the limits of a particular system is still challenging to assess, and a direct comparison between systems is very difficult. Trackability is the first measure of how different systems compare; however, it is just a start.  Its main weakness is not taking into consideration the bandwidth of the controller.  Additionally, application-specific metrics (e.g., related to fluorescence lifetime measurements) are needed to quantify how a particular system and configuration will meet different experimental aims. A selection tool that utilizes these (yet-to-be-developed) application-specific metrics to enable a selection of the RT-FD-SPT methods that best fits a particular experimental aim would be a valuable contribution to the community.

\subsection{Adoption of RT-FD-SPT}
%adoption
The final challenge in RT-FD-SPT is driving awareness and adoption of these methods.  The development of the core technologies of RT-FD-SPT is well established, with multiple techniques available for the simultaneous tracking and spectroscopic measurements of single particles and molecules in their native environments. We believe that these methods represent a valuable addition to the toolbox of researchers and that they enable a vast array of research questions to finally become addressable.  Unfortunately, there are still barriers to adoption that must be overcome to make these techniques more widespread. One important consideration is the need to be able to reproduce a specific experimental setup and a particular experiment. Currently, the descriptions of experiments in most published works are not detailed enough to be re-created in another laboratory without significant additional experimentation. In many cases, while the hardware for a method may be relatively simple, the software development is quite complex and methods for parameter tuning are often poorly described. Many systems rely on proprietary tools such as NI LabVIEW rather than open-source hardware, making it difficult to share code. A detailed protocol paper or open-source software approach would go a long way to making the field accessible to newcomers or non-specialists. Another barrier to adoption and development is the cost of the RT-FD-SPT microscope. As shown in Table~\ref{tab:cost}, many of the techniques discussed in this review are expensive to implement. Additionally, it may be unclear to a newcomer which aspects of a setup are essential or worth spending more money on and which can be achieved using a lower-cost part. While we hope this paper helps address that somewhat, often the determination of what is important depends on the specific application and experimental conditions. Finally, the inter-disciplinary nature of RT-FD-SPT can be a significant hurdle. Many of the research groups engaged in RT-FD-SPT have teams that span biology to electrical engineering. The development of open-source hardware and software will go a long way to enabling the equitable adoption of RT-FD-SPT, and enabling various applications.

To date, RT-FD-SPT methods have been applied in a handful of applications and produced excellent results (see Section~\ref{sec:Survey}). We believe that with increasing awareness and adoption, many more biologists and spectroscopists will see the great potential of the technique to studying in real time any biological process that involves the translational motion of molecules, macromolecules and/or larger particles like viruses and vesicles. The technique has the potential to revolutionize many research fields, including those of virology and drug delivery \cite{Welsher2015}. We would especially like to see more spectroscopy applications. In the SMS field, there is a large push toward single-molecule studies in natural environments, and RT-FD-SPT is perfectly suited to this task. We believe it can be a widely used technique, alongside other advanced techniques such as superresolution microscopy and nonlinear microscopy. The combination of microscopic and spectroscopic measurements is potentially a powerful avenue of research and should be of special interest. We also encourage using RT-FD-SPT to revisit thousands of earlier studies, but now for (macro)molecules or particles in live cells as they explore various environments for longer times and probing their dynamics and interactions with improved spatiotemporal resolution. 

\medskip
% Acknowledgements
\medskip
\textbf{Acknowledgements} \par %delete if not applicable))
B.v.H. and N.A.V contributed equally to this work. S.B.A. acknowledges funding from NIH through grant NIGMS 5R01GM117039. T.P.J.K. acknowledges funding from the South African National Research Foundation through grant N00500/120387.

% References
\medskip

% Use the following code if you wish to generate your bibliography with BibTeX;
% replace the string "MSP-template" below with the name(s) of
% the BibTeX data base(s) you want to use.
% The resulting bibliography-output (the content of the .bbl file)
% must be pasted back into this file before submission.
% Please also include your BibTeX data base file(s) in your submission
% so that we can re-run BibTeX if necessary.

%\textbf{References}\\
\bibliographystyle{MSP}
\bibliography{ControlsBib.bib, SPT_Review.bib}

% Figures/tables and captions
% Permission statements are required for all figures reproduced or adapted from previously published articles/sources. Please also ensure that all necessary permissions to reproduce images have been received
% Please remove these statements for original figures

% \begin{figure}
%   \includegraphics[width=\linewidth]{SmallTemplate/placeholder-image.png}
%   \caption{Figure 1 caption goes here. Reproduced with permission.\textsuperscript{[Ref.]} Copyright Year, Publisher. }
%   \label{fig:boat1}
% \end{figure}

% \begin{table}
%  \caption{Table 1 caption}
%   \begin{tabular}[htbp]{@{}lll@{}}
%     \hline
%     Description 1 & Description 2 & Description 3 \\
%     \hline
%     Row 1, Col 1  & Row 1, Col 2  & Row 1, Col 3  \\
%     Row 2, Col 1  & Row 2, Col 2  & Row 2, Col 3  \\
%     \hline
%   \end{tabular}
% \end{table}

% Please provide Biographies and photos for Essays, Feature Articles, Progress Reports, Reviews, and Perspectives for those authors who should be highlighted  
% These should be at most 100 words long
% For other article types this section can be removed
% Photographs should be 40mm broad and 50 mm high
\clearpage
\begin{figure}
  \includegraphics{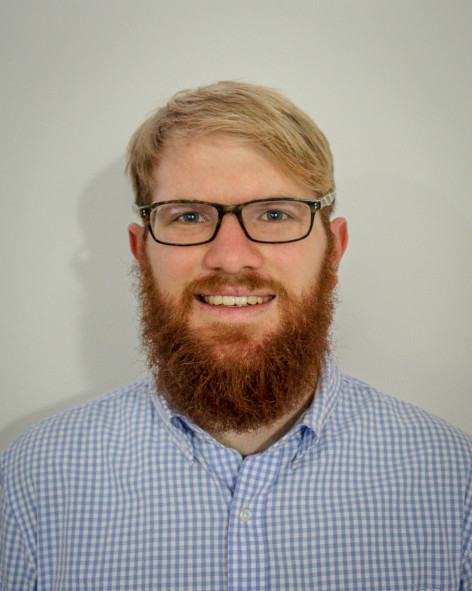}
  \caption*{Bertus van Heerden received a B.Sc., B.Sc. (Hons.) and M.Sc. in Physics from the University of Pretoria in South Africa and is now pursuing a Ph.D. at the same university. His research focuses on single molecule spectroscopy of light-harvesting complexes and single-particle tracking.}
\end{figure}

\begin{figure}
  \includegraphics{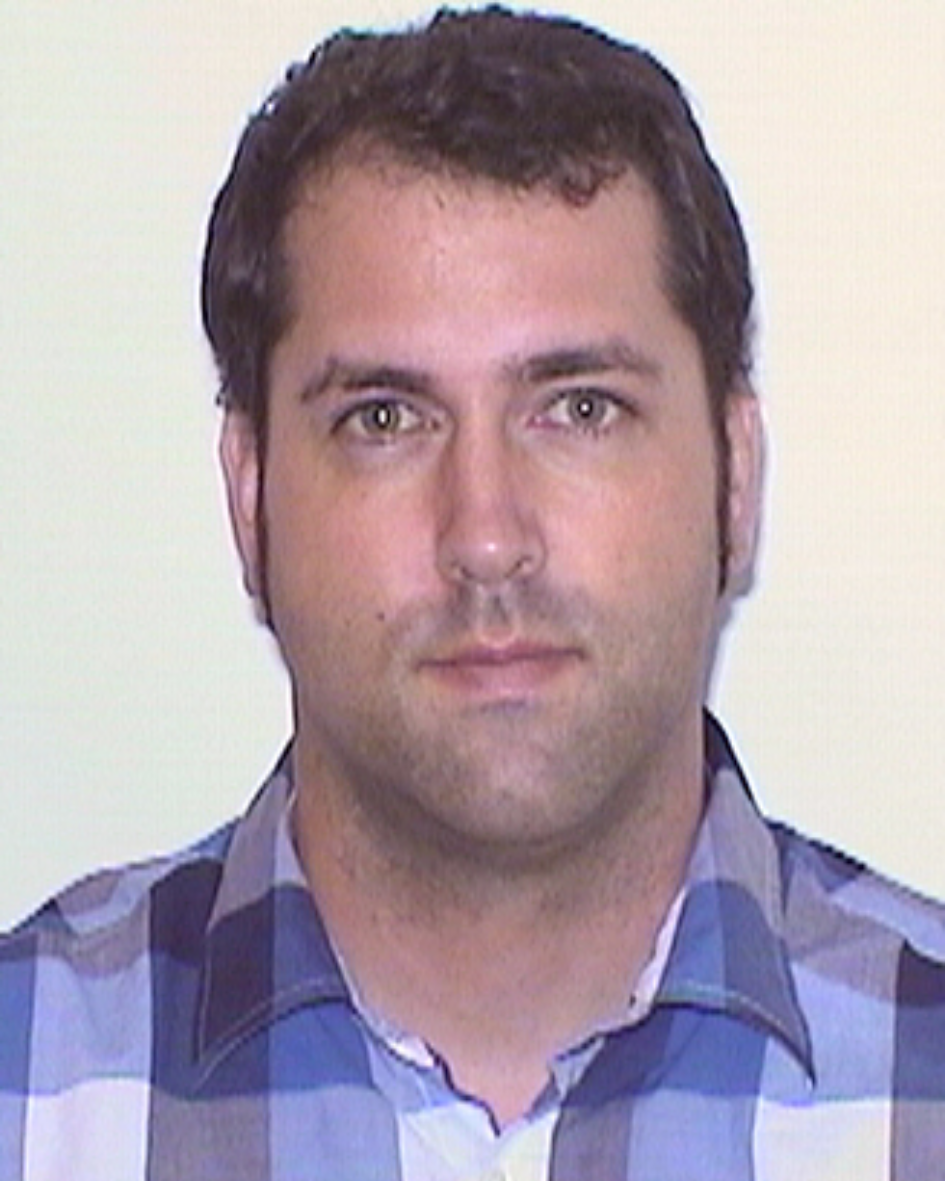}
  \caption*{Nicholas A. Vickers started his career at California Polytechnic State University where he pursued a B.S. in Materials Engineering and B.A. in Physics, specializing in microfluidics, microfabrication, and micro-optics. He went on to serve as a process engineer at Soraa, followed by Technical Accounts Manager and Field Service Engineer at ASMPT. He is currently a Mechanical Engineering PhD candidate at Boston University, Boston, MA, USA. His current research focuses on information optimal control of tracking spectroscopy microscopes. His research interests include high-speed control systems for optical microscopes, information optimization, engineered point spread functions, and single-particle tracking.}
\end{figure}

\begin{figure}
  \includegraphics{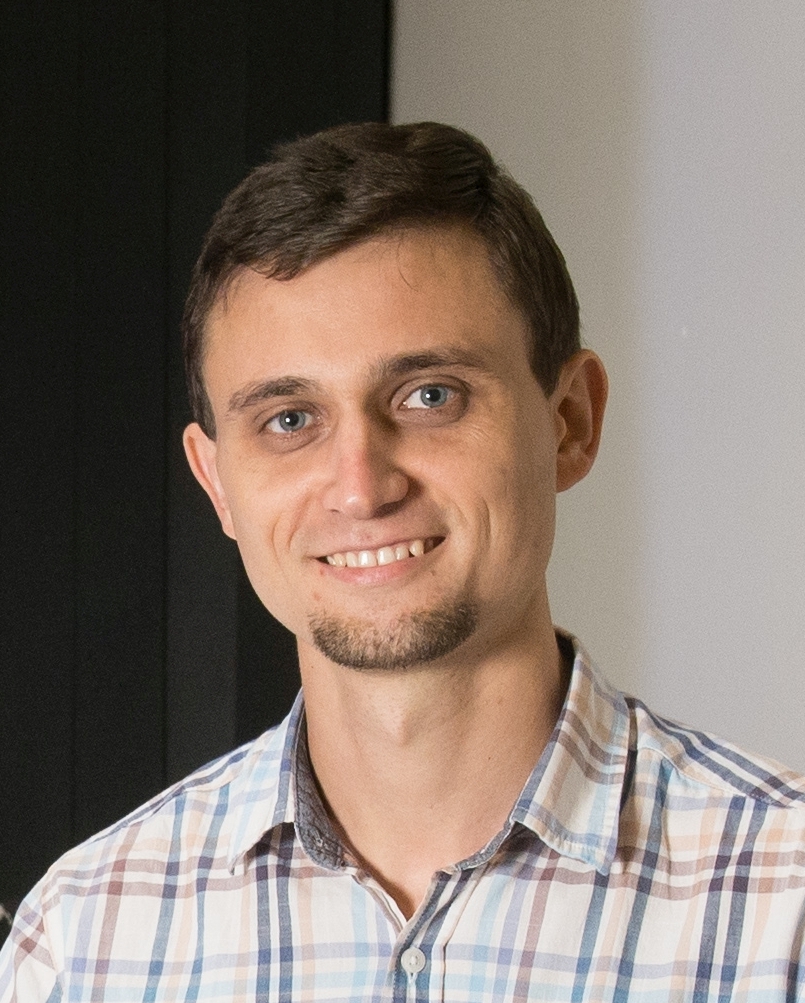}
  \caption*{Tjaart P.J. Krüger received his Ph.D. in Biophysics from the Vrije Universiteit Amsterdam in The Netherlands and thereafter moved to South Africa to establish biophysics research at the University of Pretoria. He is currently an associate professor in physics at the same university. His research focuses mainly on single molecule spectroscopy and transient absorption spectroscopy of photosynthetic light-harvesting complexes, semiconductor nanostructured materials, organic polymers, and metallic nanoparticles.}
\end{figure}

\begin{figure}
  \includegraphics{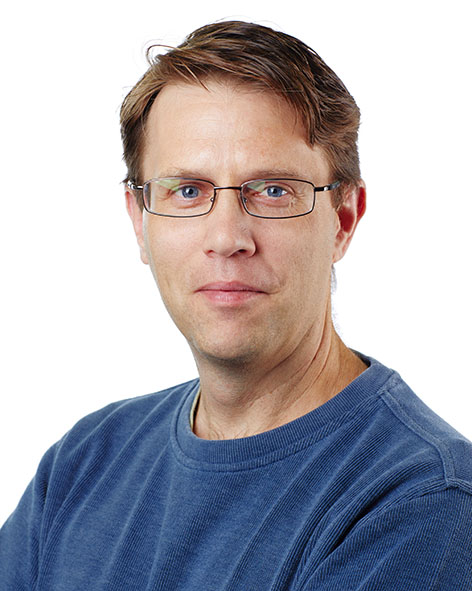}
  \caption*{Sean B. Andersson received a B.S. in engineering and applied physics from Cornell University, Ithaca, NY, USA, an M.S. in mechanical engineering from Stanford University, Stanford, CA, USA, and a Ph.D. in electrical and computer engineering from the University of Maryland, College Park, MD, USA. He is currently a professor of mechanical engineering and of systems engineering with Boston University, Boston, MA, USA. His research interests include systems and control theory with applications in single-particle tracking, biophysics, scanning probe microscopy, and robotics.}
\end{figure}

% Table of contents entry should be 50 - 60 words long
% Image should be 55 mm broad and 50 mm high or 110 mm broad and 20 mm high

% \begin{figure}
% \textbf{Table of Contents}\\
% \medskip
%   \includegraphics{Figures/ToCimage.png}
%   \medskip
%   \caption*{An emerging approach to studying freely moving single particles and molecules is Real-Time Feedback-Driven Single-Particle Tracking (RT-FD-SPT) and RT-FD-SPT spectroscopy. This paradigm enables simultaneous tracking and spectroscopic measurements of particles in their native environments. This review surveys methods and applications of RT-FD-SPT, describes the main components of existing approaches and how each influences performance, and concludes with future perspectives.}
% \end{figure}

\end{document}